\documentstyle[11pt,aaspp4]{article}

\makeatletter
\let\DOTSI\relax
\def\RIfM@{\relax\ifmmode}%
\def\FN@{\futurelet\next}%
\newcount\intno@
\def\iint{\DOTSI\intno@\tw@\FN@\ints@}%
\def\iiint{\DOTSI\intno@\thr@@\FN@\ints@}%
\def\iiiint{\DOTSI\intno@4 \FN@\ints@}%
\def\idotsint{\DOTSI\intno@\z@\FN@\ints@}%
\def\ints@{\findlimits@\ints@@}%
\newif\iflimtoken@
\newif\iflimits@
\def\findlimits@{\limtoken@true\ifx\next\limits\limits@true
 \else\ifx\next\nolimits\limits@false\else
 \limtoken@false\ifx\ilimits@\nolimits\limits@false\else
 \ifinner\limits@false\else\limits@true\fi\fi\fi\fi}%
\def\multint@{\int\ifnum\intno@=\z@\intdots@                                %1
 \else\intkern@\fi                                                          %2
 \ifnum\intno@>\tw@\int\intkern@\fi                                         %3
 \ifnum\intno@>\thr@@\int\intkern@\fi                                       %4
 \int}%                                                                     %5
\def\multintlimits@{\intop\ifnum\intno@=\z@\intdots@\else\intkern@\fi
 \ifnum\intno@>\tw@\intop\intkern@\fi
 \ifnum\intno@>\thr@@\intop\intkern@\fi\intop}%
\def\intic@{\mathchoice{\hskip.5em}{\hskip.4em}{\hskip.4em}{\hskip.4em}}%
\def\negintic@{\mathchoice
 {\hskip-.5em}{\hskip-.4em}{\hskip-.4em}{\hskip-.4em}}%
\def\ints@@{\iflimtoken@                                                    %1
 \def\ints@@@{\iflimits@\negintic@\mathop{\intic@\multintlimits@}\limits    %2
  \else\multint@\nolimits\fi                                                %3
  \eat@}%                                                                   %4
 \else                                                                      %5
 \def\ints@@@{\iflimits@\negintic@
  \mathop{\intic@\multintlimits@}\limits\else
  \multint@\nolimits\fi}\fi\ints@@@}%
\def\intkern@{\mathchoice{\!\!\!}{\!\!}{\!\!}{\!\!}}%
\def\plaincdots@{\mathinner{\cdotp\cdotp\cdotp}}%
\def\intdots@{\mathchoice{\plaincdots@}%
 {{\cdotp}\mkern1.5mu{\cdotp}\mkern1.5mu{\cdotp}}%
 {{\cdotp}\mkern1mu{\cdotp}\mkern1mu{\cdotp}}%
 {{\cdotp}\mkern1mu{\cdotp}\mkern1mu{\cdotp}}}%
\def\rmfam{\z@}%
\newif\iffirstchoice@
\firstchoice@true
\def\textfonti{\the\textfont\@ne}%
\def\textfontii{\the\textfont\tw@}%
\def\text{\RIfM@\expandafter\text@\else\expandafter\text@@\fi}%
\def\text@@#1{\leavevmode\hbox{#1}}%
\def\text@#1{\mathchoice
 {\hbox{\everymath{\displaystyle}\def\textfonti{\the\textfont\@ne}%
  \def\textfontii{\the\textfont\tw@}\textdef@@ T#1}}%
 {\hbox{\firstchoice@false
  \everymath{\textstyle}\def\textfonti{\the\textfont\@ne}%
  \def\textfontii{\the\textfont\tw@}\textdef@@ T#1}}%
 {\hbox{\firstchoice@false
  \everymath{\scriptstyle}\def\textfonti{\the\scriptfont\@ne}%
  \def\textfontii{\the\scriptfont\tw@}\textdef@@ S\rm#1}}%
 {\hbox{\firstchoice@false
  \everymath{\scriptscriptstyle}\def\textfonti
  {\the\scriptscriptfont\@ne}%
  \def\textfontii{\the\scriptscriptfont\tw@}\textdef@@ s\rm#1}}}%
\def\textdef@@#1{\textdef@#1\rm\textdef@#1\bf\textdef@#1\sl\textdef@#1\it}%
\def\DN@{\def\next@}%
\def\eat@#1{}%
\def\textdef@#1#2{%
 \DN@{\csname\expandafter\eat@\string#2fam\endcsname}%
 \if S#1\edef#2{\the\scriptfont\next@\relax}%
 \else\if s#1\edef#2{\the\scriptscriptfont\next@\relax}%
 \else\edef#2{\the\textfont\next@\relax}\fi\fi}%
\def\Let@{\relax\iffalse{\fi\let\\=\cr\iffalse}\fi}%
\def\vspace@{\def\vspace##1{\crcr\noalign{\vskip##1\relax}}}%
\def\multilimits@{\bgroup\vspace@\Let@
 \baselineskip\fontdimen10 \scriptfont\tw@
 \advance\baselineskip\fontdimen12 \scriptfont\tw@
 \lineskip\thr@@\fontdimen8 \scriptfont\thr@@
 \lineskiplimit\lineskip
 \vbox\bgroup\ialign\bgroup\hfil$\m@th\scriptstyle{##}$\hfil\crcr}%
\def\Sb{_\multilimits@}%
\def\endSb{\crcr\egroup\egroup\egroup}%
\def\Sp{^\multilimits@}%

\newdimen\ex@
\def\rightarrowfill@#1{$#1\m@th\mathord-\mkern-6mu\cleaders
 \hbox{$#1\mkern-2mu\mathord-\mkern-2mu$}\hfill
 \mkern-6mu\mathord\rightarrow$}%
\def\leftarrowfill@#1{$#1\m@th\mathord\leftarrow\mkern-6mu\cleaders
 \hbox{$#1\mkern-2mu\mathord-\mkern-2mu$}\hfill\mkern-6mu\mathord-$}%
\def\leftrightarrowfill@#1{$#1\m@th\mathord\leftarrow\mkern-6mu\cleaders
 \hbox{$#1\mkern-2mu\mathord-\mkern-2mu$}\hfill
 \mkern-6mu\mathord\rightarrow$}%
\def\overrightarrow{\mathpalette\overrightarrow@}%
\def\overrightarrow@#1#2{\vbox{\ialign{##\crcr\rightarrowfill@#1\crcr
 \noalign{\kern-\ex@\nointerlineskip}$\m@th\hfil#1#2\hfil$\crcr}}}%

\def\overleftarrow{\mathpalette\overleftarrow@}%
\def\overleftarrow@#1#2{\vbox{\ialign{##\crcr\leftarrowfill@#1\crcr
 \noalign{\kern-\ex@\nointerlineskip}$\m@th\hfil#1#2\hfil$\crcr}}}%
\def\overleftrightarrow{\mathpalette\overleftrightarrow@}%
\def\overleftrightarrow@#1#2{\vbox{\ialign{##\crcr\leftrightarrowfill@#1\crcr
 \noalign{\kern-\ex@\nointerlineskip}$\m@th\hfil#1#2\hfil$\crcr}}}%
\def\underrightarrow{\mathpalette\underrightarrow@}%
\def\underrightarrow@#1#2{\vtop{\ialign{##\crcr$\m@th\hfil#1#2\hfil$\crcr
 \noalign{\nointerlineskip}\rightarrowfill@#1\crcr}}}%

\def\underleftarrow{\mathpalette\underleftarrow@}%
\def\underleftarrow@#1#2{\vtop{\ialign{##\crcr$\m@th\hfil#1#2\hfil$\crcr
 \noalign{\nointerlineskip}\leftarrowfill@#1\crcr}}}%
\def\underleftrightarrow{\mathpalette\underleftrightarrow@}%
\def\underleftrightarrow@#1#2{\vtop{\ialign{##\crcr$\m@th\hfil#1#2\hfil$\crcr
 \noalign{\nointerlineskip}\leftrightarrowfill@#1\crcr}}}%
\newcount\GRAPHICSTYPE
\GRAPHICSTYPE=\z@
\def\GRAPHICSPS#1{%
 \ifcase\GRAPHICSTYPE%\GRAPHICSTYPE=0
  ps: #1%
 \or%\GRAPHICSTYPE=1
  language "PS", include "#1"%
 \or%\GRAPHICSTYPE=2
  #1%
 \fi
}%
\def\graffile#1#2#3#4{%
 \ifnum\GRAPHICSTYPE=\tw@
  \@ifundefined{psfig}{\input psfig.tex}{}%
  \psfig{file=#1, height=#3, width=#2}%
 \else
  \leavevmode\raise -#4 \hbox{%
   \raise #3 \hbox{\rule{0.003in}{0.003in}\special{#1}}%
   }%
  {\raise -#4 \hbox to #2 {\vrule height#3 width\z@ depth\z@\hfil}}%
 \fi
}%
\def\draftbox#1#2#3#4{%
 \leavevmode\raise -#4 \hbox{%
  \frame{\rlap{\protect\tiny #1}\hbox to #2%
   {\vrule height#3 width\z@ depth\z@\hfil}%
  }%
 }%
}%
\newcount\draft
\draft=\z@
\def\GRAPHIC#1#2#3#4#5{%
 \ifnum\draft=\@ne\draftbox{#2}{#3}{#4}{#5}%
  \else\graffile{#1}{#3}{#4}{#5}%
  \fi
 }%
\def\addtoLaTeXparams#1{\edef\LaTeXparams{\LaTeXparams #1}}%
\def\doFRAMEparams#1{\readFRAMEparams#1\end}%
\def\readFRAMEparams#1{%
 \ifx#1\end%
  \let\next=\relax
  \else
  \ifx#1i\dispkind=\z@\fi
  \ifx#1d\dispkind=\@ne\fi
  \ifx#1f\dispkind=\tw@\fi
  \ifx#1t\addtoLaTeXparams{t}\fi
  \ifx#1b\addtoLaTeXparams{b}\fi
  \ifx#1p\addtoLaTeXparams{p}\fi
  \ifx#1h\addtoLaTeXparams{h}\fi
  \let\next=\readFRAMEparams
  \fi
 \next
 }%
\def\IFRAME#1#2#3#4#5{\GRAPHIC{#5}{#4}{#1}{#2}{#3}}%
\def\DFRAME#1#2#3#4{%
 \begin{center}\GRAPHIC{#4}{#3}{#1}{#2}{\z@}\end{center}%
 }%
\def\FFRAME#1#2#3#4#5#6#7{%
 \begin{figure}[#1]%
  \begin{center}\GRAPHIC{#7}{#6}{#2}{#3}{\z@}\end{center}%
  \caption{\label{#5}#4}%
  \end{figure}%
 }%
\newcount\dispkind%
\def\FRAME#1#2#3#4#5#6#7#8{%
 \def\LaTeXparams{}%
 \dispkind=\z@
 \def\LaTeXparams{}%
 \doFRAMEparams{#1}%
 \ifnum\dispkind=\z@\IFRAME{#2}{#3}{#4}{#7}{#8}\else
  \ifnum\dispkind=\@ne\DFRAME{#2}{#3}{#7}{#8}\else
   \ifnum\dispkind=\tw@
    \edef\@tempa{\noexpand\FFRAME{\LaTeXparams}}%
    \@tempa{#2}{#3}{#5}{#6}{#7}{#8}%
    \fi
   \fi
  \fi
 }%
\long\def\QQQ#1#2{\long\expandafter\def\csname#1\endcsname{#2}}%
\def\QTP#1{}%
\long\def\QQA#1#2{}%
\def\QTR#1#2{{\csname#1\endcsname #2}}%(gp) Is this the best?
\def\EXPAND#1[#2]#3{}%
\def\NOEXPAND#1[#2]#3{}%
\def\LaTeXparent#1{}%
\def\QTagDef#1#2#3{}%
\def\QQfnmark#1{\footnotemark}

\@ifundefined{INDEX}{\def\INDEX#1#2{}{}}{}%
\@ifundefined{SUBINDEX}{\def\SUBINDEX#1#2#3{}{}{}}{}%
\def\initial#1{\bigbreak{\raggedright\large\bf #1}\kern 2\p@\penalty3000}%
\@ifundefined{abstract}{%
 \def\abstract{%
  \if@twocolumn
   \section*{Abstract (Not appropriate in this style!)}%
   \else \small 
   \begin{center}{\bf Abstract\vspace{-.5em}\vspace{\z@}}\end{center}%
   \quotation 
   \fi
  }%
 }{%
 }%
\@ifundefined{endabstract}{\def\endabstract
  {\if@twocolumn\else\endquotation\fi}}{}%
\@ifundefined{maketitle}{\def\maketitle#1{}}{}%
\@ifundefined{affiliation}{\def\affiliation#1{}}{}%
\@ifundefined{proof}{}{}%
\@ifundefined{endproof}{}{}%
\@ifundefined{newfield}{\def\newfield#1#2{}}{}%
\@ifundefined{chapter}{\def\chapter#1{\par(Chapter head:)#1\par }%
 \newcount\c@chapter}{}%
\@ifundefined{part}{\def\part#1{\par(Part head:)#1\par }}{}%
\@ifundefined{section}{\def\section#1{\par(Section head:)#1\par }}{}%
\@ifundefined{subsection}{\def\subsection#1%
 {\par(Subsection head:)#1\par }}{}%
\@ifundefined{subsubsection}{\def\subsubsection#1%
 {\par(Subsubsection head:)#1\par }}{}%
\@ifundefined{paragraph}{\def\paragraph#1%
 {\par(Subsubsubsection head:)#1\par }}{}%
\@ifundefined{subparagraph}{\def\subparagraph#1%
 {\par(Subsubsubsubsection head:)#1\par }}{}%
\@ifundefined{therefore}{}{}%
\@ifundefined{backepsilon}{}{}%
\@ifundefined{yen}{}{}%
\@ifundefined{registered}{\def\registered{\relax\ifmmode{}\r@gistered
                                                \else$\m@th\r@gistered$\fi}%
 \def\r@gistered{^{\ooalign
  {\hfil\raise.07ex\hbox{$\scriptstyle\rm\text{R}$}\hfil\crcr
  \mathhexbox20D}}}}{}%
\@ifundefined{Eth}{}{}%
\@ifundefined{eth}{}{}%
\@ifundefined{Thorn}{}{}%
\@ifundefined{thorn}{}{}%
\@ifundefined{degree}{}{}%
\def\BibTeX{{\rm B\kern-.05em{\sc i\kern-.025em b}\kern-.08em
    T\kern-.1667em\lower.7ex\hbox{E}\kern-.125emX}}%
\newdimen\theight
\def\Column{%
 \vadjust{\setbox\z@=\hbox{\scriptsize\quad\quad tcol}%
  \theight=\ht\z@\advance\theight by \dp\z@\advance\theight by \lineskip
  \kern -\theight \vbox to \theight{%
   \rightline{\rlap{\box\z@}}%
   \vss
   }%
  }%
 }%
\def\qed{%
 \ifhmode\unskip\nobreak\fi\ifmmode\ifinner\else\hskip5\p@\fi\fi
 \hbox{\hskip5\p@\vrule width4\p@ height6\p@ depth1.5\p@\hskip\p@}%
 }%
\def\miss{\hbox{\vrule height2\p@ width 2\p@ depth\z@}}%
\def\tcol#1{{\baselineskip=6\p@ \vcenter{#1}} \Column}  %
\makeatother

\begin{document}

\singlespace

\title{On the Dynamics and Structure of Three-Dimensional Trans-Alfv\'enic Jets}

\author{Philip E. Hardee and Alexander Rosen}  
\affil{Department of Physics \& Astronomy \\ The University of Alabama \\
Tuscaloosa, AL 35487 \\ hardee@athena.astr.ua.edu, rosen@eclipse.astr.ua.edu}

\begin{abstract}

Three-dimensional magnetohydrodynamical simulations of strongly
magnetized ``light'' conical jets have been performed. An investigation
of the transition from sub-Alfv\'enic to super-Alfv\'enic flow has been
made for nearly poloidal and for helical magnetic fields. The  jets are
stable to asymmetric modes of jet distortion provided they are
sub-Alfv\'enic over most of their interior but destabilize rapidly when
they become on average super-Alfv\'enic.  The jets are precessed at the
origin and the resulting small amplitude azimuthal motion is
communicated down the jet to the Alfv\'en point where it couples to a
slowly moving and rapidly growing helical twist.  Significant jet
rotation can contribute to destabilization via increase in the velocity
shear between the jet and the external medium.  Destabilization is
accompanied by significant mass entrainment and the jets slow down
significantly as denser external material is entrained.  Synchrotron
intensity images satisfactorily reveal large scale helical structures
but have trouble distinguishing a large amplitude elliptical jet
distortion that appears as an apparent pinching  in an intensity
image.  Smaller scale jet distortions are not clearly revealed in
intensity images, largely as a result of the relatively small total
pressure variations that accompany destabilization and growing
distortions.  Fractional polarization is high as a result of the strong
ordered magnetic fields except where the intensity image suggests
cancellation of polarization vectors by integration through twisted
structures.

\end{abstract}


\section{Introduction}

Highly collimated outflows are observed to emanate from the centers of
galaxies and quasars, from neutron star and black hole binary star
systems, and from protostellar systems. Hydrodynamic jet models can
account for many aspects of the dynamics and morphology of the extended
jets -- both galactic and extragalactic. However, simple flux
conservation arguments imply and recent jet acceleration and
collimation schemes require dynamically strong magnetic fields close to
the central engine. Numerical studies, e.g., Meier, Payne, \& Lind
(1996); Ouyed, Pudritz, \& Stone (1997); Ouyed \& Pudritz (1997);
Romanova et al.\ (1997), show that the jets created in this fashion
pass through slow magnetosonic, Alfv\'enic, and fast magnetosonic
critical points.  The ultimate jet velocity may depend on the
configuration of the magnetic field (Meier et al.\ 1997), and the jets
accelerate up to asymptotic speeds that may be only a few times the Alfv\'en
speed at the Alfv\'en point -- at the Alfv\`en point the jet speed
equals the Alfv\`en speed (Camenzind 1997). This basic acceleration and
collimation process may be the same for all classes of objects that
emit jets (Livio 1997).

Highly collimated flows are susceptible to Kelvin-Helmholtz (K-H) or
current driven (CD) instabilities. The K-H instability of
three-dimensional (3D) jets with purely poloidal or purely toroidal
magnetic fields (Ray 1981; Ferrari, Trussoni, \& Zaninetti 1981;
Fiedler \& Jones 1984; Bodo et al.\ 1989), and of jets containing
force-free helical magnetic fields (Appl \& Camenzind 1992; Appl 1996)
has been extensively investigated.  Additional investigations have
considered the potential role of current driven pinching of toroidally
magnetized columns (Begelman 1998). At least for force-free helical
magnetic fields it appears that the K-H instability exhibits faster
growth and is more likely to be responsible for producing asymmetric
structure than current driven instability (Appl 1996). In general,
spatial or temporal growth rates associated with the K-H instability
are found to increase as the magnetosonic Mach number decreases
provided the jet is super-Alfv\'enic.  Unlike purely fluid flows which
are unstable when subsonic, the poloidally magnetized jet is predicted
to be nearly completely stabilized to the K-H instability when the jet
is sub-Alfv\'enic. Additionally, an appropriately configured dynamically
significant magnetic field may have a stabilizing influence on the
super-Alfv\'enic jet.  Nevertheless, the fact that magnetically
accelerated and collimated jets must pass through a super-Alfv\'enic
but transmagnetosonic region implies a potential zone of enhanced
instability just downstream of the Alfv\'en point.

Previous numerical work designed to investigate the effect of strong
magnetic fields on jet stability (Hardee et al.\ 1992; Hardee \& Clarke
1995) were conducted using two dimensional (2D) slab jet geometry. A
slab jet is spatially resolved along two Cartesian axes and is
effectively infinite in extent in the third dimension. The 2D nature of
such simulations reduced computer memory and CPU requirements. A
theoretical analysis of the stability properties of the slab jet
reveals that the jet is K-H unstable to a symmetric pinching mode and
an asymmetric sinusoidal mode that provide reasonable analogs to the
pinching and helical twisting of a 3D jet. The numerical simulations
confirmed that the stability properties of the axially magnetized slab
jet behaved according to predictions made by a stability analysis. In
particular, it was demonstrated that a super-Alfv\'enic jet becomes
more stable as the magnetosonic Mach number increases with a
destabilization length varying approximately proportional to the
magnetosonic Mach number. The slab jet simulations showed a complete
stabilization of the jet to sinusoidal distortion when the jet was
sub-Alfv\'enic, and showed the predicted rapid destabilization at the
Alfv\'en point where the flow becomes super-Alfv\'enic but is
transmagnetosonic. The numerical simulations also showed that magnetic
tension can significantly modify the development of instability in the
nonlinear regime and in 2D prevent disruption of the flow. However, the
2D slab jet has no analog to the higher order modes of distortion
(elliptical, triangular, rectangular, etc.) of the 3D jet. These modes
make the 3D jet more unstable than the 2D slab jet and lead to enhanced
spatial mass entrainment rates (cf., Rosen et al.\ 1999, hereafter
RHCJ). Momentum and mass exchange with an external environment more
dense than the jet results in relatively rapid loss of the initial high
collimation, and the outwards flow broadens and slows as denser
external material is heated, mixed with, and accelerated by the lighter
jet fluid. Thus, it is of considerable interest to see the effect of
very strong magnetic fields on mass entrainment in 3D.  Additionally, it is
of interest to ascertain the types of jet structures arising near the
Alfv\'en point, and to search for a connection between these structures
and observed jet structures.

It is our purpose here to begin a numerical investigation of the
dynamical and stability properties of strongly magnetized flows in 3D.
Ultimately we hope to shed some light on the structures associated with
magnetized jet configurations near the jet acceleration and
collimation region, and to provide some connection between the
acceleration and collimation region and observed jet structures on
larger spatial scales.  In this paper we analyze results from 3D
simulations designed to study the predicted rapid destabilization at
the transition between sub-Alfv\'enic and super-Alfv\'enic flow and to
study the effect of strong fields on mass entrainment. In \S 2 the
numerical setup and results of the numerical simulations are
presented.  The simulations are initialized by establishing a
cylindrical helically magnetized jet across a computational grid in an
unmagnetized surrounding medium with pressure gradient devised to
result in a constant expansion of the jet once pressure equilibrium is
achieved. Thus, these simulations are relevant to astrophysical jets
far behind the propagating jet head and in an assumed quasi-steady
state region. Total synchrotron intensity images and fractional
polarization vectors provide a connection between jet dynamics and
potentially observable jet structures in extragalactic jets, and also
provide some insight into potential structures in radiatively cooled
protostellar or Seyfert jets. In \S 3 the structures expected to arise
as a result of the K-H instability are presented and compared to
structures appearing in the numerical simulations. Finally, in \S 4 we
summarize our results and discuss some of the implications for
astrophysical jets.

\section{Numerical Simulations}

\subsection{Initialization}

Simulations were performed using the three dimensional MHD code
ZEUS-3D, an Eulerian finite-difference code using the Consistent Method
of Characteristics (CMoC) which solves the transverse momentum
transport and magnetic induction equations simultaneously and in a {\it
planar split} fashion (Clarke 1996). Interpolations were carried out by
a second-order accurate monotonic upwinded time-centered scheme (van
Leer 1977) and a von-Neumann Richtmyer artificial viscosity was used to
stabilize shocks. The code has been thoroughly tested via MHD test
suites as described by Stone et al.\ (1992) and Clarke (1996) to
establish the reliability of the techniques.

All simulations are initialized by establishing a cylindrical jet
across a 3D Cartesian grid resolved into 130 $\times $ 130 $\times $
370 zones. With this grid the simulations required about 126 Mwords of
memory on the Cray C90 at the Pittsburgh Supercomputer Center. Thirty
uniform zones span the initial jet diameter, $2R_0$, along the
transverse Cartesian axes ($x$-axis and $y$-axis). Outside the uniform
grid zones, the zones are ratioed where each subsequent zone
increases in size by a factor 1.05. Altogether the 130 zones along the
transverse Cartesian axes span a total distance of $30R_0$. Along the
$z$-axis 280 uniform zones span a distance of $40R_0$ outwards from the
jet origin. An additional 90 ratioed zones span an additional distance
of $40R_0$ where each subsequent zone increases in size by a factor
1.02. Altogether the 370 zones along the $z$-axis span a total distance
of $80R_0$. Outflow boundary conditions are used except where the jet
enters the grid where inflow boundary conditions are used. The use of a
non-uniform grid such as we are employing has been shown to have the
beneficial effect of reducing reflections off the grid boundaries as a
result of increased dissipation of disturbances (Bodo et al.\ 1995).

The jets  are initialized across the computational grid with a uniform
density $\rho _{jt}$ and initial radius $R_0$. The magnetic field in
the jet is initialized with a uniform axial component, $B_z$, and a
toroidal magnetic component with functional form $B_\phi =B_\phi
^{pk}\sin ^2[\pi f(r)]$ where for $r<r_{pk}$, $f(r)=0.5(r/r_{pk})^a$,
and for $r_{pk}<r<r_{\max }$ , $ f(r)=1.0-0.5\left[ \left( 1-r/r_{\max
}\right) /(1-r_{pk}/r_{\max })\right] ^b$. In these simulations the
toroidal component increases to a maximum, $ B_\phi ^{pk}$, at
$r_{pk}=0.5R_0$, and declines to zero at $r_{\max }=0.9R_0 $ so that
initially all currents flow within the jet. This particular toroidal
profile is not physically motivated but with $a=b=0.315$ provides a
broad cross sectional region within the jet in which the toroidal
magnetic component is relatively constant (nearly constant helical
pitch, plasma beta, and magnetosonic speed) and also is well behaved
numerically. In the external medium the magnetic field is equal to
zero. The equation of hydromagnetic equilibrium
$$
\frac d{dr}\left( p_{jt}(r)+\frac{B_z^2(r)}{8\pi }+\frac{B_\phi ^2(r)}{8\pi}
\right) =-\frac{B_\phi ^2(r)}{4\pi r}\text{ ,} 
$$
where the term on the right hand side describes the effects of magnetic
tension, has been used to establish a suitable radial gas pressure
profile in the jet by varying the jet temperature. The sonic,
Alfv\'enic and magnetosonic Mach numbers in the jet are
$M_{jt}(r)\equiv u/a_{jt}(r)$, $ M_A(r)\equiv u/V_A(r)$ and
$M_{ms}(r)\equiv u/a_{ms}(r)$ where $ a_{jt}^2(r)=\Gamma p(r)/\rho
_{jt}$ , $\Gamma =5/3$, $V_A^2(r)=B^2(r)/4\pi \rho _{jt}$, and we
define a jet magnetosonic speed as $a_{ms}(r)\equiv
(a_{jt}^2+V_A^2)^{1/2}$. We categorize simulations by the magnetosonic
Mach number on the jet axis but note that the dynamics can depend on
details of the magnetic and temperature profiles. Since internal
dynamics and timescales involve wave propagation across a jet with
sound, Alfv\'en and magnetosonic speeds which are a function of jet
radius, we can define radial averages as, for example,
$$
\left\langle M_{ms}\right\rangle \equiv \frac
1{R_{jt}}\int_0^{R_{jt}}M_{ms}(r)dr 
$$
that will differ for different magnetic, density and temperature
profiles.

In the simulations the external medium is isothermal and the external
density, $\rho _{ex}(z)$, declines to produce a pressure gradient, 
$p_{ex}(z)\propto \rho _{ex}(z)$, that is designed to lead to a constant
expansion, $R_{jt}(z)=(1+z/80R_0)R_0$, of a constant velocity adiabatic
jet containing uniform poloidal magnetic field and an internal toroidal
magnetic field that provides some confinement, i.e.,
$$
\rho _{ex}(z)=\frac{\left[ (R_{jt}/R_0)^{-10/3}+C_p(R_{jt}/R_0)^{-4}-C_\phi
(R_{jt}/R_0)^{-2}\right] }{\left[ 1+C_p-C_\phi \right] }\rho _{ex}(0) \text{ .}
$$
The values of $C_p$ and $C_\phi$ depend on the poloidal and
toroidal field strengths, and the ratio of the magnetic pressure
relative to the thermal pressure. The jet speed is initialized so that
the jets are sub-Alfv\'enic initially, but after expansion to achieve
pressure equilibrium with the external medium cross the Alfv\'en and
fast magnetosonic points on the computational grid. All jets are
initialized with a velocity $u=4a_{ex}$ and a density $\rho
_{jt}=0.029\rho _{ex}(0)$. The jet sound, Alfv\'en, and magnetosonic
speeds at jet center normalized to the external sound speed; radial
averages of the jet sonic, Alfv\'enic, and magnetosonic Mach numbers at
the inlet, and values of $C_p$ and $C_\phi$ for the simulations are
given in Table 1. Absolute values for jet speed, density, temperature,
and magnetic field strength are completely determined by choosing
values for the external density, $\rho _{ex}(0)$, and the external
temperature, $T_{ex}$, or sound speed, $a_{ex}$. An absolute length
scale is determined by choosing a value for $R_0$.

Simulations A and B contain a primarily poloidal magnetic field with
only a weak toroidal component, $B_\phi ^{pk}/B_z\sim $ (A) $0.086$ \&
(B) $0.05$ at the inlet, so as to be directly comparable to predictions
made by a linear stability analysis (\S 3). In addition, jet parameters
in simulation A at the inlet are very nearly identical to jet
parameters in a 2D MHD expanding slab jet simulation (Hardee \& Clarke
1995). In simulation B the poloidal and toroidal magnetic field
strengths are reduced and the jet thermal pressure is increased
relative to simulation A. In simulation B  the Alfv\'en and
magnetosonic points move closer to the inlet.  Simulations C \& D
contain a poloidal magnetic field that is the same as in simulation B
but now the toroidal component $B_\phi ^{pk}/B_z\sim 0.44$ at the
inlet. The thermal pressure and toroidal magnetic field radial profiles
used in the four simulations are shown in Figure 1. The total magnetic
field strength in simulations C \& D has been designed to be comparable
to the total magnetic field strength in simulation A at an axial
distance of $z=40R_0$.  If constant adiabatic expansion and velocity,
along with decline in the jet density and poloidal magnetic field
proportional to $R_{jt}^{-2}$ are assumed, then the Alfv\'en point
evaluated on the jet axis where the toroidal component of the magnetic
field is zero would occur at axial distances of (A) $40R_0$ and (B, C \&
D) $20R_0$.

In all simulations the jet is driven by a periodic precession of the
jet velocity, $u$, at an angle of 0.01 radian relative to the $z$-axis with
an angular frequency $\omega =0.5a_{ex}/R_0$. The initial transverse
motion imparted to the jet by this precession is well within the linear
regime. The precession serves to break the symmetry and the
precessional frequency is chosen to be below the theoretically
predicted maximally unstable frequency associated with helical twisting
of a K-H unstable supermagnetosonic jet with jet radius $R_{jt}\geq
R_0$. In simulations A, B and C the precession is in a counterclockwise
sense when viewed outwards from the inlet. This direction of precession
induces a helical twist in the same sense as that of the magnetic field
helicity and helical wavefronts are at shallow angles to the helically
twisted magnetic field lines. In simulation D the precession is
clockwise and helical wavefronts associated with the precessional
motion will be at larger angles relative to the helically twisted
magnetic field lines.

\subsection{Simulation Results}

In all simulations the jets expand rapidly, and after about five
dynamical times, $\tau _d\equiv (a_{ex}/R_0)t\approx 5$, have achieved
an approximate static pressure equilibrium with the surrounding
medium.  In all cases the jets achieve a nearly constant expansion rate
on the computational grid with $R_{jt}\approx 2R_0$ when $z=80R_0$.
After dynamical times $\tau _d=$ (A) 68, (B) 56, and (C \& D) 44 the
numerical simulations have reached a quasi-steady state out to axial
distances between $45R_0$ and $60R_0$ depending on the simulation. The
dynamical times correspond to $\sim 3.5$ precessional periods and $\sim
3.5$ flow through times through an axial distance of $50R_0$ in
simulations C \& D and correspondingly more precessional periods and
flow through times in simulations A \& B. The simulations were
terminated before a quasi-steady state was achieved across the entire
computational grid because excessive CPU time would have been
required.  The simulations required (A) 200, (B) 160, and (C \& D) 360
CPU hours on the Cray C90 at the Pittsburgh Supercomputer Center.

Plots of the velocity components along the $z$-axis shown in Figure 2
reveal that the jets accelerate in response to the magnetic and thermal
pressure gradients. In simulations A \& B with primarily poloidal
magnetic fields the jets accelerate up to some asymptotic speed early
in the simulation but do not quite reach the initial asymptotic speed
at later dynamical times as they are slowed by the onset of
instability.  Instability is manifested by the rapid growth in the
amplitude and by the oscillation of the transverse velocity components,
and by the fluctuation and decline in the axial velocity. Significant
fluctuation in axial and transverse velocity components begins at an
axial distance of (A) $\sim 40R_0$ and (B) $\sim 20R_0$.  In
simulations A \& B transverse velocities are less than 2\% of the jet
speed inside the destabilization point but grow to values as large as
(A) 35\% and (B) 40\% of the jet speed shortly after destabilization.
Transverse velocity oscillations with scale lengths $ \lambda _h\sim $
(A) $7.6R_0$ and (B) $7.1R_0$ are out of phase in the orthogonal
transverse directions suggesting a 3D ``helical'' distortion, and not,
for example, a 2D ``sinusoidal'' distortion of the jet beam.
Fluctuations in the axial velocity are the result of the jet flow being
displaced off the $z$-axis.

More acceleration is evident in simulations C \& D than in simulation
B.  Recall that the poloidal magnetic field is the same in simulations
B, C \& D. Nevertheless simulations C \& D show the same basic initial
behavior as simulations A \& B, i.e., jet acceleration up to some
asymptotic speed early in the simulation followed by development of
instability at later dynamical times. Instability occurs somewhat
closer to the inlet in simulations C \& D than in simulation B. Note an
indication of relatively large transverse velocities immediately
outside the inlet, up to 10\% of the jet speed near to the jet axis,
which decrease but then become as large as 35\% of the jet speed as the
jets destabilize. The initial large transverse motions are indicative
of jet rotation in these two simulations, and occur as the code
modifies the initial input static equilibrium state to an appropriate
dynamic equilibrium state, i.e., $\nabla \times ({\bf u}\times {\bf
B})=0$.  We note that dynamic equilibrium is achieved across
the computational grid as the jets expand to achieve static pressure
balance with the external medium. We also note that dynamic equilibrium
is achieved  throughout the duration of the simulation within one jet
radius of the inlet. The jet rotational speed is on the order of the
external sound speed and is about 25\% of the initial jet speed.  In
these two simulations this azimuthal motion is vastly larger than the
azimuthal motion induced at the inlet by the precession. Simulations C
\& D show significant fluctuation in axial and transverse velocity
components associated with instability beginning at an axial distance
of $ \sim 15R_0$. Major transverse velocity oscillations with scale
length $ \lambda _h$ $\sim 6.5R_0$ (not too different from simulation
B) are complicated by other smaller scale features but indicate helical
motion of the jet beam. In simulations C \& D more rapid and larger
amplitude fluctuation is seen in the axial velocity than in simulations
A \& B.

Plots of the axial velocity along with the sonic, Alfv\'enic, and
magnetosonic speeds along the $z$-axis including the Alfv\'en and fast
magnetosonic points and plots of transverse structure along the
$x$-axis between the Alfv\'en and fast magnetosonic points are shown in
Figure 3. Jet acceleration downstream of the inlet results in Alfv\'en
and fast magnetosonic points on the $z$-axis at distances $z_{{\rm A}
}\sim $ (A) $37R_0$, (B) $8R_0$, (C) $6R_0$, and (D) $8R_0$, and $z_{
{\rm ms}} \sim$ (A) $42R_0$, and (B, C \& D) $20R_0-24R_0$,
respectively.  The choice of a cut point just downstream of the
Alfv\'en point on the jet axis reveals the relatively undisturbed jet
profile before significant instability appears.  The transverse profile
is particularly important as the jet may still remain sub-Alfv\'enic
off the axis and stable. The plots of transverse structure show that
the poloidally magnetized jets in simulations A \& B have relatively
flat (top hat) density, temperature, and magnetic field strength and
velocity profiles, while the helically magnetized jets in simulations C
\& D show considerably more transverse structure. The jets can become
super-Alfv\'enic off the jet axis at a significantly different point
than on the jet axis, and, for example, the average Alfv\'enic Mach
number $\left\langle M_{{\rm A}}\right\rangle \approx 1$ at an axial
distance $\left\langle z_{{\rm A}}\right\rangle \sim $ (A) $ 35R_0$,
(B) $8R_0$, and (C \& D) $14R_0$.  In simulation A a small enhancement
in the Alfv\'en speed on the jet axis leaves the jet sub-Alfv\'enic on
the axis while already super-Alfv\'enic off the axis.  The helically
magnetized jets in simulations C \& D become super-Alfv\'enic on the
jet axis and at the jet surface where the toroidal component of the
magnetic field is zero while remaining sub-Alfv\'enic in much of the
jet interior. Note that in simulations C \& D  the Alfv\'en speed in
the jet interior is as much as 9\% higher than on the jet axis but the
total jet speed including rotation is only about 3\% higher than on the
jet axis. In all simulations the jets remain stable to large scale
velocity fluctuations at the jet center until they become on average
super-Alfv\'enic.  Significant large scale velocity fluctuations
develop before the jet can become supermagnetosonic.

The mass entrained by the jets, and the average velocity of jet plus
``entrained'' material in the four simulations are plotted in Figure
4.  In particular, we define the mass per unit length, $\sigma (z)$, at
any point along the jet as $\sigma =\int_Af\rho dydx$, where $A$ is the
cross sectional area of the computational domain at axial position $z$,
and $f$ is set to 1 if the local magnetic field is above 4\% of the
expected maximum field strength along the jet at $z$ [$B(z) > 0.04
B_{jt,max}(z)$], and $f$ is set to 0 otherwise. We define the entrained
mass by the presence of a magnetic field since only the jet material is
initially magnetized. The setting of $f$\ to 1 or 0 effectively assumes
that zones with any fraction of jet material, as defined by the
presence of magnetic field, are considered mixed with the external
medium in that zone. This technique provides results similar to
estimating mass entrainment by using a threshold axial velocity
(RHCJ).  Setting the switch at a magnetic field strength of 4\% of the
expected maximum strength at $z$ reduces the effects of numerical
diffusion, and also reduces the sensitivity of the value of $\sigma $
to a small diffusion of the field into the external medium which is
much denser than the jet material. Even so there is a large increase in
$\sigma $ at the jet inlet.  Simulations A and B provide a useful
baseline value of $\sigma /\sigma _{jt} \approx 10$ ($\sigma _{jt}$ is
the expected value at the inlet) that could be the result of numerical
diffusion.  Thus, for example we might assume that a value of $\sigma
/\sigma _{jt}=20$ infers an entrained mass equal to the ``initial'' jet
mass per unit length. We note that the high density in the external
medium relative to the jet density at the inlet results in the observed
value of $\sigma$ at the inlet if the jet magnetic field promptly
diffuses radially by about two computational zones.  Note that the
average velocity of ``entrained'' plus jet material is very low at the
inlet in simulations A and B.  If we also consider Figure 5 which shows
gray scale axial velocity cross sections, we infer that dense material
``entrained'' near to the inlet is moving very slowly in a thin sheath
around a rapidly moving jet core.  The slow motion of sheath material,
typically more than an order of magnitude less than the core speed,
means that insufficient time has passed in the simulation for the
majority of ``entrained '' material to have moved far downstream and
most mass has been picked up relatively locally.  For example, the
noticeable slow decrease in ``entrained'' mass in simulation A out to
$z \approx 42R_{jt}$ occurs because the jet density at larger distance
is higher relative to the external density -- recall the external
medium is isothermal and in equilibrium with an expanding
adiabatic magnetized jet, thus the external density falls faster than the
expanding jet's density -- with result that mass at larger distance
falling within the $B(z) > 0.04 B_{jt,max}(z)$ criterion is a smaller
fraction of the jet's mass.

In simulations A \& B the onset of significant mass entrainment occurs
relatively abruptly at axial distances of (A) $42R_0$ and (B) $19R_0$
as large transverse velocities develop, and the average velocity of jet
plus ``entrained'' material shows a significant decline beyond these
points. The maximum value of the mass per unit length is $(\sigma
/\sigma _{jt})^{\max }\approx $ (A) $14$ and (B) $45 $.  These values
would imply an entrained mass of about 0.4 and 3.5 times the initial
jet mass per unit length, respectively, if we assume $\sigma /\sigma
_{jt} = 10$ is an appropriate baseline level.  In simulations C \& D
mass entrainment begins at an axial distance of $\sim 6R_0$, possibly
slightly before the jet becomes super-Alfv\'enic on average when only
the axial velocity is considered.  This mass entrainment begins even
before significant transverse velocity fluctuations are apparent on the
jet axis. In simulations C \& D the maximum value of the mass per unit
length is $(\sigma /\sigma _{jt})^{\max }\approx 55$.  This value
implies an entrained mass of about 4.5 times the initial jet mass per
unit length if we assume $\sigma /\sigma _{jt} = 10$ is an appropriate
baseline level. While an accurate quantitative value for the amount of
mass entrained cannot be determined from these simulations, we can
conclude that significant mass entrainment occurs when the jets become
super-Alfv\'enic on average.  We have examined our simulations to see
if significant forward momentum flux, $\sigma v_z^2$, is carried by
``unmixed '' external material, i.e., material with $B(z) < 0.04
B_{jt,max}(z)$. We evaluate the momentum flux in the simulations for
$15 < z/R_0 < 60$. At least approximately these minimum and maximum
limits on $z$ span the range beyond the initial jet acceleration region
out to the quasi-steady state limit.  In simulations A \& B the
``unmixed'' material in this range carries $\lesssim 2\%$ \& $\lesssim
5\%$ of the momentum flux, respectively. We note that acceleration has
led to a total momentum flux in this range about 1.15 and 1.35 times
higher than at the inlet in simulations A \& B, respectively. In
simulations C \& D the ``unmixed'' material between $15 < z/R_0 < 30$
carries $\lesssim 6\%$ of the momentum flux and acceleration has led to
a total momentum flux in this range about 1.25 times higher than at the
inlet.   From $30 < z/R_0 < 60$ in simulations  C \& D the ``unmixed''
material carries $\lesssim 15\%$ of the momentum flux and the total
momentum flux is now only about 70\% of that at the inlet.  The
reduction in total momentum flux in simulations C \& D compared to
simulation B at $15 < z/R_0 < 30$ is likely the result of an elliptical
distortion that appears in simulations C \& D at small $z$. While
momentum imparted to the ``unmixed'' material is not insignificant, it
is clear that the majority of the jet kinetic energy remains carried by
the ``mixed'' material.

In the super-Alfv\'enic mass entraining region we note that decline in
the mass entrained at large axial distances is a consequence of
termination of the simulations before a quasi-steady state is achieved
across the entire computational grid.  Decline in the entrained mass
and increase in the velocity of jet plus entrained mass at large
distances suggests that simulations A \& B have reached a quasi-steady
state out to about $55R_0-60R_0$, and simulations C \& D out to about
$45R_0$. We note that the spatial mass entrainment rate after
destabilization is less in simulation A relative to B. The spatial mass
entrainment rate in simulations C \& D is higher  than in simulation B
for $z \leq 25R_0$ but the entrained mass in simulation B becomes
comparable to C \& D for $25R_0 \leq z \leq 35R_0$. The entrained mass
in simulation B continues to increase steadily out to $z \approx 40R_0$.
In simulations C \& D the entrained mass remains relatively constant
for $25R_0 \leq z \leq 39R_0$ with a sudden jump in the mass entrained
at $z\sim 39R_0$. The entrained mass in simulations B, C \& D is
comparable between $40R_0-50R_0$ and remains relatively constant over
this interval. This may indicate saturation like that found in other
simulations (Bodo et al.\ 1995; RHCJ) or in this case be an effect of
insufficient computational and flow through time.

Jet axial velocity cross sections shown in Figure 5 illustrate
development of the surface distortions that promote mixing and mass
entrainment, and that can move the jet flow completely off the initial
axis. Small scale surface distortions in simulations A \& B have formed
before the jets become super-Alfv\'enic on the axis.  These jets become
super-Alfv\'enic near to the jet surface first, partly as a result of
prompt entrainment at the jet inlet which raises the jet density,
lowers the magnetic field strength, and reduces the Alfv\'en speed near
the jet surface. Clearly these small scale surface distortions do not
lead to significant mass entrainment. In particular we note a
``rectangular'' distortion in simulation A appearing in the panel at
$z=24R_0$ with an admixture of ``elliptical'' and ``triangular''
distortion evident in the panels at $z=48R_0$ and $54R_0$. The
``rectangular'' distortion rotates through $90^{\arcdeg}$ in an axial
distance of $\lambda _r\sim 18R_0$. Note that jet distortions in
simulation A are still confined near to the jet surface even at axial
distances of $50R_0$ and the entrained mass remains low. The jet in
simulation B exhibits considerable structure that consists of a
combination of helical, elliptical, triangular, and rectangular
distortions. For example, the panels at $z=18R_0$ and $24R_0$ show
evidence for triangular and rectangular distortion plus helical
displacement, and  primarily triangular distortion plus helical
displacement, respectively, and an elliptical distortion is apparent in
the panel at $z=42R_0$. The development of significant jet distortion
at $z=24R_0 $ is coincident with significant mass entrainment. Note
that the jet flow has moved completely off the $z$-axis in the panel at
$z=48R_0$ as a result of helical displacement. Shifting of the jet off
the $z$-axis leads to the decrease in axial and transverse velocities
seen in Figure 2 at large axial distances in simulation B.  In both of
these simulations the jets indicate a diminishing higher speed core
within a growing more slowly moving sheath at large distances.

The jet axial velocity cross section show that simulations C \& D
develop similarly, but with significantly different behavior from
simulations A \& B. The cross sections show a surface elliptical
distortion by an axial distance of $6R_0$ with a scale length for
rotation through $180^{\arcdeg}$ of $\lambda _e\sim 8R_0$.  The higher
order surface corrugations that appeared in simulations A \& B are
suppressed in these simulations.  Ultimately these jets appear to
become hollow in the axial velocity cross section at axial distances of
$\sim 40R_0$.  This occurs close to the location of the sudden jump in
entrained mass in these two simulations. In these two simulations we
almost lose the higher speed jet core entirely at large distances.  The
gray scale cross sections indicate that significant momentum is carried
by a comparably large transverse region in simulations B, C \& D.  We
will consider jet cross section distortion along with accompanying
pressure and velocity fluctuation in more detail in the next section.

In Figure 6 total synchrotron intensity images containing fractional
polarization B-vectors (vectors indicating the magnetic field
direction) formed by line-of-sight integrations through the
computational domain reveal the types of observational structures that
might appear downstream of the Alfv\'en point. To some extent these
images also reveal the extent of jet spreading as only the jet fluid is
magnetized. To generate these images a synchrotron emissivity is
defined by $p_{jt}(B\sin \theta )^{3/2}$ where $ \theta $ is the angle
made by the magnetic field with respect to the line of sight, and the
simulated intensity and fractional polarization B-vectors are formed
from the Stokes parameters. This emissivity mimics synchrotron emission
from a system in which the energy and number densities of the
relativistic particles are proportional to the energy and number
densities of the thermal fluid. This simplistic assumption is necessary
when the relativistic particles are not explicitly tracked (Clarke,
Norman, \& Burns 1989). While not designed to be comparable to line
emission images from radiatively cooling Seyfert or protostellar jets,
at least approximately these images would be representative of emission
from a gas with radiative cooling proportional to $n^{2.5}T$, i.e., the
jet density and poloidal magnetic field ($\sin \theta \approx 1$) are
initially constant across the jet and we assume that compressions of
the magnetic field are comparable to compressions in the particle
number density ($B\propto n$).

Patterns seen in the intensity image for simulation A are not readily
identifiable with the prominent rectangular jet distortions apparent in
the velocity cross sections although evidence of structure is apparent
in the intensity image.  The helical twisting indicated by the
transverse velocities in Figure 2 produces only a barely discernible
sinusoidal oscillation in the intensity image. The intensity image for
simulation B readily reveals a sinusoidal oscillation that is
identifiable with the helical motion of the jet indicated by the
transverse velocities in Figure 2. Rotating elliptical distortion is
not readily identifiable in the intensity image although some narrowing
and broadening in the intensity image between $30R_0-50R_0$ is
associated with this distortion. The elliptical distortion seen in the
velocity cross sections in simulations C \& D is apparent in the
intensity images as an oscillation in jet width between axial distances
of $10R_0-20R_0 $. It is important to note that the oscillation in jet
width in total intensity is not the result of axisymmetric pinching.
The intensity images for simulations C \& D show an unusual irregularly
oscillating structure between $30R_0-40R_0$ which terminates as the
axial velocity cross sections indicate development of a hollow jet.
This intensity structure indicates that the jet beam develops an
irregular helically twisted structure which terminates in a
circumferential loop in simulation C, and which appears to loop back on
itself slightly in simulation D at axial distances between $
42R_0-43R_0$. In both these simulations the twist is in a
counterclockwise sense when viewed downstream from the jet inlet (in
the same sense as jet rotation). The polarization B-vectors overlaid on
the intensity images are in general aligned with twisted structures in
the intensity images, and the fractional polarization is relatively
large except where the intensity image suggests cancellation of
B-vectors by integration through twisted structures, e.g., note the
much smaller B-vectors associated with the loop structure in
simulations C \& D at axial distances of $42R_0-43R_0$. The high
fractional polarization and preferred orientation along the flow
direction is a consequence of the strong ordered poloidal magnetic
field component. Note that the more helical field in simulations C \& D
would still give polarization B-vectors initially aligned with the jet
axis.

In the next section we investigate the types of structures that should
develop in the transition between sub-Alfv\'enic and super-Alfv\'enic flow,
compare theoretically predicted structures to those observed in the
numerical simulations, and discuss the relationship between structure
observed in the intensity images and the underlying flow dynamics.

\section{Jet Structure}

\subsection{Stability Theory}

The stability of an axially magnetized cylindrical jet with top hat
profile residing in a uniform unmagnetized medium has been investigated
in a number of papers, e.g., Ray 1981, Ferrari, Trussoni, \& Zaninetti
1981; Bodo et al.\ 1989, and we briefly review and extend the results
here. It is assumed that the jet is a cylinder of radius $R$, having a
uniform density, $\rho _{jt}$, a uniform internal axial magnetic field,
$B_{jt}$, and a uniform velocity, $ u $. Inclusion of a small
non-uniform toroidal magnetic field component like that used in
simulations A \& B will not significantly modify results obtained from
an analysis incorporating only axial magnetic fields, although we
expect significant effects associated with the much stronger toroidal
magnetic field used in simulations C \& D, cf., Appl \& Camenzind
(1992), Appl (1996), RHCJ. The external medium is
assumed to have a uniform density, $\rho _{ex}$, and to contain no
magnetic field. The jet is established in static total pressure balance
with the external medium where the total static uniform pressure is
$p_{jt}^{*}\equiv p_{jt}+B_{jt}^2/8\pi = $ $p_{ex}^{*}=p_{ex}$. The
relatively slow jet expansion in the numerical simulations is not
expected to significantly modify results based on a completely
uniform external medium (Hardee 1984). In cylindrical geometry a random
perturbation of $\rho _1$, ${\bf u}_{1\text{,}}$ $p_1$, and ${\bf B} _1
$ to an initial equilibrium state $\rho _0$, ${\bf u}_{0\text{,}}$
$p_0$, and ${\bf B}_0$ can be considered to consist of Fourier
components of the form
\begin{equation}
\eqnum{1}f_1(r,\phi ,z)=f_1(r)\exp [i(kz\pm n\phi -\omega t)] 
\end{equation}
where flow is along the $z$-axis, and $r$ is in the radial
direction with the flow bounded by $r=R$. In cylindrical geometry $k$
is the longitudinal wavenumber, $n$ is an integer azimuthal wavenumber,
for $n>0$ the wavefronts are at an angle to the flow direction, the
angle of the wavevector relative to the flow direction is $\theta =\tan
(n/kR)$, and $+n$ and $-n$ refer to wave propagation in the clockwise
and counterclockwise sense, respectively, when viewed outwards along
the flow direction. In equation (1) $n=$ 0, 1, 2, 3, 4, etc. correspond
to pinching, helical, elliptical, triangular, rectangular, etc. normal
mode distortions of the jet, respectively.  Propagation and growth or
damping of the Fourier components is described by a dispersion
relation [cf., Hardee, Clarke, \& Rosen (1997) eq.(A6), hereafter HCR].

In general, each normal mode, $n$, contains a single ``surface'' wave
and multiple ``body'' wave solutions that satisfy the dispersion
relation. The behavior of the solutions can be investigated
analytically in the limit $\omega \rightarrow 0$.

In this limit the real part of the pinch mode ($n=0$) surface
wave solution becomes (Hardee 1995)
\begin{equation}
\eqnum{2}\frac \omega k\approx u\pm \left\{ \frac 12\left( V_A^2+\frac{
V_A^2a_{jt}^2}{a_{ms}{}^2}\right) \pm \frac 12\left[ \left( V_{A,jt}^2+\frac{
V_{A,jt}^2a_{jt}^2}{a_{ms}{}^2}\right) ^2-4\frac{V_A^4a_{jt}^2}{a_{ms}{}^2}
\right] ^{1/2}\right\} ^{1/2}\text{ .} 
\end{equation}
The imaginary part of the solution is vanishingly small in the low
frequency limit. These solutions are related to fast ($+$) and slow
($-$) magnetosonic waves propagating with ($u+$) and against ($u-$) the
jet flow speed $u$, but strongly modified by the jet-external medium
interface. Numerical solution of the dispersion relation reveals that a
growing solution is associated with the backwards moving (in the jet
fluid reference frame) solution related to the slow magnetosonic wave
and the pinch mode surface wave is unstable on sub-Alfv\'enic and
super-Alfv\'enic jets. 

When a jet is sub-Alfv\'enic the helical and higher order ($n>0$)
surface modes are stable (Bodo et al.\ 1989; Hardee et al.\ 1992) and
have an outwards moving purely real solution given by (e.g., Hardee \&
Clarke 1995)
\begin{equation}
\eqnum{3}\omega /k\approx u+V_A\text{ .} 
\end{equation}
On the supermagnetosonic jet all higher order modes ($n>0$) have
surface wave solutions given by
\begin{equation}
\eqnum{4a}\frac{\omega}{k} \approx \frac {\eta}{1+\eta}u  
\left\{ 1\pm i
\frac{\left[ 1 - (1+\eta)V_A^2/u^2\right]^{1/2}}{\eta^{1/2}}
\right\}
\text{ ,}
\end{equation}
or
\begin{equation}
\eqnum{4b}\frac{ku}\omega \approx \frac 1{1-V_A^2/u^2} 
\left\{ 1\pm i
\frac{\left[ 1 - (1+\eta)V_A^2/u^2\right]^{1/2}}{\eta^{1/2}}
\right\}
 \text{ ,} 
\end{equation}
where the density ratio $\eta \equiv \rho _{jt}/\rho _{ex}$. Spatial
growth corresponds to the minus sign in equation (4b) and a negative
value for the imaginary part of the complex wavenumber. Note that in
the dense jet limit, i.e., $\eta \rightarrow \infty $, equations (4a)
and (4b) reduce to equation (3) with $ \omega /k\approx u\pm V_A$, and
thus the surface waves are related to Alfv\'en waves propagating with
and against the jet flow speed but strongly modified by the
jet-external medium interface. The unstable growing solution is
associated with the backwards moving (in the jet fluid reference frame)
wave. Equation (4a) indicates a surface wave speed in the observer
frame, $(\omega /k)_{Real}\approx [\eta /(1+\eta )]u$, that is a strong
function of the density ratio. When a jet is super-Alfv\'enic but
transmagnetosonic the propagation speed and the growth rate at higher
frequency can only be determined by numerical solution of the
dispersion relation.

In the low frequency limit purely real body wave solutions are given by
\begin{equation}
\eqnum{5}kR\approx \frac{(n+2m-1/2)\pi /2}{\left[
\{M_{ms}^2/[1-(M_{ms}/M_{jt}M_A)^2]\}-1\right] ^{1/2}} 
\end{equation}
where $m\geq 1$ is an integer. However, unstable body wave solutions
exist only when the denominator in equation (5) is real. This occurs if
the jet speed is slightly below the slow magnetosonic speed, $u <
v_{ms}^s$, and $u > a_{jt}V_A/(a_{jt}^2+V_A^2)^{1/2}$, or if the jet
speed is above the fast magnetosonic speed, $u > v_{ms}^f$, (Bodo et
al.\ 1989; Hardee et al.\ 1992) where for wavevectors parallel to the
axial magnetic field $v_{ms}^s=$ Min$(a_{jt},V_A)$ and $v_{ms}^f=$ Max$
(a_{jt},V_A)$. These body wave solutions are growing at higher
frequencies.

Displacements, {\boldmath $\xi$}$(r,\phi ,z)$, of jet fluid from an
initial position $(r,\phi ,z)$ are given by  {\boldmath $\xi$}$(r,\phi
,z)=$ {\boldmath $\xi$}$(r)\exp [i(kz \pm n\phi -\omega t)]$ with the
{\boldmath $\xi$}$(r,\phi ,z)$ given by equations (A9) in HCR, and
where $\omega $ and $k$ are normal mode solutions to the dispersion
relation. In general {\boldmath $\xi$}$(r)$ is complex and
displacements, {\boldmath $\xi$}$(r,\phi ,z)$, the accompanying
velocity perturbation, ${\bf u}_1$, and the total pressure
perturbation, $p_1^{*}\equiv p_1+{\bf B}_1{\bf \cdot B} _0/4\pi$, can
be written in the form
\begin{equation}
\eqnum{6a}\text{\boldmath $\xi$}(r,\phi _s,z_s)={\bf A}(r)e^{i{\bf \Delta }(r)}\xi
_{r,n}^s\exp [i(kz_s\pm n\phi _s-\omega t)]\text{ ,} 
\end{equation}
\begin{equation}
\eqnum{6b}{\bf u}_1=-i(\omega -ku){\bf A}(r)e^{i{\bf \Delta }(r)}\xi
_{r,n}^s\exp [i(kz_s\pm n\phi _s-\omega t)]\text{ ,} 
\end{equation}
\begin{equation}
\eqnum{6c}p_{1jt}^{*}(r)=B(r)e^{i\Delta_p (r)}\xi _{r,n}^s\exp [i(kz_s\pm
n\phi _s-\omega t)]\text{ ,} 
\end{equation}
where $\xi _{r,n}^s\equiv \xi _{r,n}(R)$, $\phi _s$ and $z_s$ are now
the radial displacement, azimuthal angle and axial position at the jet
surface.  

Fluid displacements are modified in amplitude and rotated in azimuthal
angle or shifted along the jet axis relative to those at the jet
surface by ${\bf A}(r)e^{i{\bf \Delta }(r)}$ [see eqs.(A10) in HCR].
The accompanying velocity perturbation is given by ${\bf u}_1(r,\phi
,z)=d${\boldmath $\xi$}$/dt$, and the velocity perturbation is modified
in amplitude and shifted in phase by the factor $-i(\omega -ku){\bf
A}(r)e^{i{\bf \Delta }(r)}$.

In equation (6c) $B(r)e^{i\Delta_p (r)}=(\chi _{jt}/\beta
_{jt})[J_n(\beta _{jt}r)/J_n^{\prime }(\beta _{jt}R)]$ [cf., Hardee et
al.\ (1998) eq.\ (14)] where $J_{n}$ is a Bessel function, the prime
denotes a derivative of the Bessel function with respect to its
argument,
$$
\beta _{jt}=\left[ -k^2+\frac{(\omega -ku)^4}{(\omega
-ku)^2(a_{jt}^2+V_A^2)-k^2V_A^2a_{jt}^2}\right] ^{1/2}\text{, }
$$
and 
$$
\chi _{jt}=\rho _{jt}[(\omega -ku)^2-k^2V_A^2]\text{ . }
$$

If the dependence of the radial fluid displacement inside the jet on
rotation in azimuth is small then the radial fluid displacement of a
surface wave mode $n>0$ is approximately given by $ \xi
_{r,n}(r)\approx \xi _{r,n}^s(r/R)^{n-1}$ (Hardee 1983). In general,
simulations indicate a somewhat faster fall off in amplitude relative
to the surface amplitude than that predicted by the analytical
approximation (Hardee, Clarke, \& Howell 1995). The accompanying
velocity and pressure variations produced by higher order surface modes
also are predicted to show a rapid decrease inwards.  At a constant
azimuth the body waves show a reversal in fluid displacement at null
surfaces interior to the jet surface but we do not consider them
further here as they are not important on the transmagnetosonic jet.

The stability of axially magnetized rotating supermagnetosonic jets has
been investigated by Bodo et al.\ (1996). In general, Bodo et al.\ (1996)
found that jet rotation provides some stabilization of the helical
surface mode. The stabilizing effects of rotation are more pronounced
at smaller longitudinal wavenumbers and diminish as the longitudinal
wavenumber increases. The lesser effect at higher longitudinal
wavenumbers is a consequence of the wavevector (the wavevector is at a
large angle relative to the axial velocity for small longitudinal
wavenumbers when $n>0$) becoming more aligned with the axial velocity,
hence reducing the effect of jet rotation on the velocity shear
parallel to the wavevector. However, Bodo et al.\ (1996) also found that
the helical mode corotating with jet rotation, $-n$ in eq. (1) for
counterclockwise jet rotation, is less stabilized by rotation than the
counter rotating helical mode, and that this effect is somewhat more
pronounced for stronger magnetic fields.  We expect that these results also
apply to the higher order $n>1$ surface modes which are essentially
harmonics of the $n=1$ helical mode.  Thus,  we expect modes
destabilized at the Alfv\'en point to be corotating with the jet
rotation if jet rotation dominates the precessional motion.

\subsection{Theory and Simulations Compared}

We have solved the dispersion relation numerically using root finding
techniques over a wide range of frequencies for parameters appropriate
to numerical simulations A \& B for the surface and a
representative sample of the body waves associated with the pinch
($n=0$), helical ($n=1$), elliptical ($n=2$), triangular ($n=3$) and
rectangular ($n=4 $) modes. Results for parameters appropriate to
simulation B in the sub-Alfv\'enic stable region at $z=7R_0$, in the
super-Alfv\'enic destabilization region at $z=10.3R_0$, and in the
transmagnetosonic region at $z=18.1R_0$ are shown in Figure 7.
Comparable results are found for parameters appropriate to simulation
A.  The predicted precessional perturbation frequency in the range  $
z=7R_0-18.1R_0$ is $\omega R_{jt}/u\approx 0.11$ where $\omega
R_0/a_{ex}=0.5$, $R_{jt} > R_0$, and $ u>4a_{ex}$ reflect jet expansion
and acceleration. In general, a jet with top hat profile is unstable to
the surface pinching mode when sub-Alfv\'enic but is stable to
body pinching and surface and body helical and higher order modes of
jet distortion. Similar K-H instability behavior is predicted to occur
in simulations A \& B, but with the different axial location of the
Alfv\'en point there are different growth rates and wavelengths
predicted to accompany the perturbing frequency. Simulations C \& D
have a poloidal field component identical to that in simulation B, and
differences between simulations C \& D and the theoretically predicted
unstable modes associated with simulation B will be the result of the
stronger toroidal field, different jet thermal pressure profile,
different axial location of the Alfv\'en point, and jet rotation.

The sub-Alfv\'enic solutions shown in Figure 7 indicate that equations
(2) and (3) provide excellent estimates of the wave speed for pinch and
higher order surface modes, respectively.  Note that the theory
indicates that a pinching perturbation propagates down the jet at less
than the jet speed while a precessional perturbation propagates down
the sub-Alfv\'enic jet at greater than the jet speed.
Our simulation setup provides no source for a pinching perturbation
other than minor symmetric disturbance as the jet enters the
computational domain at the inlet, and we do not expect to see pinching
even though the mode is unstable. For example, in simulation B the
e-folding (growth) length of a pinch perturbation is $\ell _e > 5R_{jt}$,
too long to provide sufficient amplification by the Alfv\'en point. We
note that pinching or helical motion corresponding to the precessional
frequency of $\omega R_{jt}/u\approx 0.11$ at $z=7R_0$ in simulation B
would result in pinching and helical wavelengths of $\lambda _p\approx
25R_{jt}$ and $\lambda _h\approx 115R_{jt}$, and the corresponding
pinch e-folding length is $\ell _e\approx 100R_{jt}$ (note
$R_{jt}\approx 1.1R_0$). The long helical wavelength and the absence of
damping implies that the initial precessional motion is communicated
nearly rigidly to the Alfv\'en point at $z$ $\sim 8R_0 $ in simulation
B. In simulation A with $\left\langle \omega R_{jt}/u\right\rangle
\approx 0.14$ between the origin and an Alfv\'en point at $\lesssim
35R_0$, the helical wavelength is comparably long, although somewhat
shorter relative to the average jet radius, i.e., $\lambda _h\approx
90\left\langle R_{jt}\right\rangle$. In simulation A there should be
significant phase lag between precession at the origin and at the
Alfv\'en point, and if the initial precession had significant amplitude
the jet would exhibit noticeable curvature.

Immediately outside the Alfv\'en point the helical and higher order
surface waves are rapidly growing, and, for example, helical e-folding
lengths are $\ell _e\sim $ (A) $3R_{jt}\approx 4.5R_0$ and (B)$
\;2R_{jt}\approx 2.2R_0$ at frequencies comparable to the precession
frequency.  The higher order surface modes grow more rapidly. The longer
predicted e-folding length for growth of helical motion in simulation A
when compared to simulation B appears in the simulations as slower
spatial development of transverse velocity oscillations in simulation A
(see Fig. 2). We note that the wave speed associated with the surface
modes at the precession frequency, $(\omega /k)_{Real}\approx [\eta
/(1+\eta )]u$, is no more than about 5\% of the jet speed. In
simulations A \& B the jet speed on the super-Alv\`enic jet is $u >
v_{ms}^f=$ Max$(a_{jt},V_A)$ and the body waves are unstable. However,
the body wave growth rates are found to be less than the pinch surface mode growth rate.

Beyond the Alfv\'en point the flow rapidly evolves into a
``transmagnetosonic'' regime where numerical solution of the dispersion
relation shows that the helical and higher order surface modes have
growth rates that are much larger than the growth rate of the surface
pinch mode at all frequencies, and now also show a distinct maximum in
the growth rates. The maximum growth rates are much larger and at much
higher frequencies than the growth rates at the precession frequency.
With the exception of the first pinch body wave which has a maximum
growth rate comparable to that of the asymmetric surface wave modes,
all body wave modes have growth rates that remain less than the growth
rate of the pinch surface wave mode and should be unimportant. At axial
distances beyond the Alfv\'en point simulations A \& B remain in the
``transmagnetosonic'' regime with only modest changes to the growth
rates and frequency--wavelength behavior. Thus, there is potential for
very rapidly developing very short (relative to the jet radius)
wavelength structure downstream of the Alfv\'en point. 

The present results provide good evidence for coupling between the
initial precession frequency, and the helical motion that develops
downstream from the Alfv\'en point in simulations A \& B even though
the initial transverse perturbation is at the 1\% level. To see that
this is so we note that the observed helical wavelengths of (A)
$7.6R_0$ and (B) $7.1R_0$ can be shown to correspond to (A) $\omega
R_{jt}/u\approx 0.10$ \& $0.22$ at $z=38.7R_0$ \& $46.6R_0$,
respectively, and (B) $\omega R_{jt}/u\approx 0.06$ \& $0.14$ at
$z=10.3R_0$ \& $18.1R_0$, respectively. These locations bracket the
region in which the precessional frequency of (A) $\omega
R_{jt}/u\approx 0.17$ at $z=40R_0$ and (B) $\omega R_{jt}/u\approx
0.11$ at $z=15R_0$ can couple to a growing helical distortion with the
observed wavelength. The rectangular surface mode that appears in the
velocity cross sections in simulation A before the jet as a whole
becomes super-Alfv\'enic provides evidence that surface modes can grow
provided the jet has a sufficiently thick super-Alfv\'enic surface
layer. The higher order surface wave modes that are readily apparent in
the velocity cross sections in simulation B (see Fig.  4) do not have
well defined wavelengths but are consistent with higher order modes
having wavelengths appropriate to the precessional frequency.

In the ``transmagnetosonic'' region in simulations A \& B the
precession frequency is about an order of magnitude less than the
frequency at which the growth rate is a maximum and the induced
wavelengths are much longer than that associated with the maximum
growth rate. The theory suggests that these ``long'' wavelength
structures can achieve relatively large amplitudes with only modest
pressure gradients because jet material does not need to move
transverse to the flow direction at a large fraction of the
magnetosonic speed, which is comparable to the flow speed. In Figure 8
we show possible maximum helical, elliptical, triangular, and
rectangular distortions appropriate to simulation B, and with
wavelengths appropriate to the precessional frequency in the
transmagnetosonic region. The maximum displacement is estimated from
the distortions evident in velocity cross sections (Fig. 5) or from the
transverse velocity oscillation (Fig. 2). The results shown in Figure 8
have used the dispersion relation solutions shown in Figure 7 in the
transmagnetosonic region at $z=18.1R_0$ and assumed a frequency of
$\omega R_{jt}/u=0.15$ to yield wavelengths representative of
wavelengths in the transmagnetosonic region at larger axial distances.
The wavelengths accompanying this frequency are $\lambda _1=5.6R_{jt}$,
$\lambda _2=5.5R_{jt} $, $\lambda _3=6.2R_{jt}$, and $\lambda
_4=8.7R_{jt}$ and the surface displacement amplitudes shown in Figure 8
are $\xi _{r,1}^s/R_{jt}=0.3$, $ \xi _{r,2}^s/R_{jt}=0.3$, $\xi
_{r,3}^s/R_{jt}=$ $0.2$, and $\xi _{r,4}^s/R_{jt}=0.15$. These
wavelengths and distortions are representative of wavelengths and
maximum distortions seen in the transmagnetosonic region.  Note that
these wavelengths correspond to 360$^{\arcdeg}$, 180$^{\arcdeg}$, 120
$^{\arcdeg}$, and 90$^{\arcdeg}$ azimuthal rotations of the helical,
elliptical, triangular, and rectangular distortions, respectively. The
total pressure variation along with axial and transverse velocity
components accompanying the distortions shown in Figure 8 are evaluated where these quantities show the largest
fluctuation,
at a radial location on the $y$-axis that is inside the maximally
deflected jet surface.

The total pressure, axial velocity and transverse velocity components
observed in simulation B along the $z$-axis and parallel to the
$z$-axis at $x = 0.7R_0$ are shown in Figure 9.  We see that the
observed amplitudes of the total pressure fluctuations and the velocity
fluctuations are comparable to the theoretical fluctuations expected to
accompany the observed displacements. Along the $z$-axis pressure and
axial velocity fluctuations are smaller than off the axis, and more
regular transverse velocity oscillation is observed along the $z$-axis
than off the axis.  This is the expected result if multiple surface
modes are operating in conjunction with a dominant helical mode. We
note that no total pressure or velocity fluctuation is predicted to
occur along the $z$-axis for elliptical and higher order modes, unless
there is significant displacement of the jet beam off the $z$-axis
associated with helical motion. On the other hand, significant pressure
and velocity fluctuation produced by higher order modes should be
observed off the $z$-axis even in the absence of significant helical
displacement. While the elliptical surface mode produces significant
displacements, pressures, and transverse motions in approximately the
outer 3/4 of the jet, the triangular and rectangular modes affect
significantly only the outer 1/2 and outer 1/3 of the jet, respectively
-- examine the displacement contours in Figure 8 -- and so only
minimally effect transverse velocities measured on the $z$-axis unless
helical motion displaces the jet off the $z$-axis by at least half the
jet radius.  Note that if multiple wave modes are operating, the total
pressure and accompanying velocity fluctuations can be as large as a
linear sum of the individual mode fluctuations. This fact allows us to
immediately rule out, for example, a combination of large amplitude
helical twist and large amplitude elliptical distortion in simulation B
because the resulting transverse velocities would be much larger than
those seen in simulation B. The relatively small total pressure
variations predicted and observed to accompany the observed jet
distortions and velocity fluctuations explain why little structure is
seen in the intensity images in Figure 6, and why only sinusoidal
oscillation associated with helical twisting and jet width fluctuation
associated with extreme elliptical distortion are evident.

The relatively strong toroidal magnetic field and accompanying large
jet rotation in simulations C \& D precludes a direct comparison to
predictions made by linear stability theory although the normal modes
of jet distortion and basic stability criteria do not change, i.e., the
jet should be stable to asymmetric (helical etc.) normal modes provided
the velocity shear parallel to the wavevector is less than the Alfv\'en
speed parallel to the wavevector (Hardee et al.\ 1992). The large jet
rotation that is present in simulations C \& D will modify the velocity
shear parallel to the wavevector of helical and higher order normal
modes at small longitudinal wavenumbers. The transverse velocity vector
plots in Figure 10 indicate that jet rotation has a significant effect
on normal mode development.  Recall that the jet rotation speed is
on the order of 20\% of the typical jet speed, and is more than an
order of magnitude larger than the azimuthal motion induced by jet
precession at the inlet.  This large rotation speed is responsible for
rapid development of the elliptical jet distortion that is apparent in
the panel in Figure 10 at $z = 4R_0$, and that can be seen in the axial
velocity cross sections (Fig. 5) and in the narrowing and broadening of
the intensity images (Fig. 6) at somewhat larger axial distance. The
elliptical distortion present in simulations C \& D corotates with jet
rotation as would be expected given that the jet rotation speed is much
larger than the azimuthal speed introduced by the precessional motion
at the inlet. The detail differences in spatial rotation of this
elliptical distortion between simulations C \& D (see Fig. 5) are the
result of the precessional motion in simulation C that corotates with
jet rotation, and in simulation D opposes jet rotation.  Rapid
appearance of the elliptical distortion in these two simulations by $z
= 4R_0$ suggests that jet rotation has increased the velocity shear
parallel to the elliptical distortion wavevector to super-Alfv\'enic
levels in a suitably deep surface layer of the jet before the jet
becomes super-Alfv\'enic on the jet axis at $z = z_{{\rm A}}\sim $
(C) $6R_0$, and (D) $8R_0$.

At axial distances beyond about $20R_0$ jet motion is dominated by a
helical twist but with an irregular wavelength that decreases up to the
loop structure seen in the intensity images (Fig. 6) at an axial
distance of $\sim 42R_0$. Beyond this distance the jets in simulations
C \& D show a longer wavelength helical distortion. Material at these
larger distances is moving somewhat faster than the loop structure.
Temporal animations of simulations C \& D suggest that material at
larger distance might move ahead of the loop structure at larger times,
that the flow might no longer be continuous across the computational
grid, and that the loop structure might develop into a slowly moving
jet front. We note, however, that magnetic tension could prevent this
development and much longer duration simulations are needed to address
the long term flow behavior. The relationship between flow dynamics and
the appearance of the loop in the intensity images is particularly
interesting and is revealed in part in Figure 11 which shows contours
of the axial velocity along with velocity vectors and contours of the
flow angle with respect to the $z$-axis in a transverse slice plane at
the location of the loop in simulation C. This figure shows that
azimuthal velocities are up to a factor of two larger than axial
velocities in the loop structure, and that there is some backwards
axial flow at this location. The large azimuthal velocities combined
with the lower axial velocities and study of animations show that the
flow is counterclockwise and around the loop structure seen in the
intensity images in simulations C \& D. However, the loop structure
itself is moving downstream at about the velocity indicated in the
axial velocity contours in Figure 11. Thus, the jet fluid flows
azimuthally around the loop but does not flow through the loop in the
downstream direction. In simulation C the bottom of the loop seen in
the intensity image (lower left quadrant in Figure 11) is moving
downstream more rapidly than the top portion of the loop. Further
evolution of the loop structure in simulation C should result in a loop
like that seen in simulation D where the bottom of the loop has
overtaken the top and the jet appears to loop back on
itself.

Previous work has indicated that development of the K-H instability in
the linear growth regime leads to a slow linear growth in the entrained
mass with subsequent rapid growth in the entrained mass in the
non-linear regime followed by saturation in the amount of entrained
mass (RHCJ). Elliptical jet distortion and subsequent filamentation
proved to be particularly destructive to jet propagation through
promotion of mass entrainment whereas higher order smaller amplitude
surface distortions had a much lesser effect. These previous findings
are reflected in the mass entrainment observed in simulations A \& B.
The lack of development of large scale distortion on the computational
grid in simulation A, a result of longer growth lengths, appears
directly related to the reduction in entrained mass relative to
simulation B. Simulation B shows considerable development of the
helical and elliptical distortions that previously have been observed
to promote mass entrainment. Simulations C \& D provide a more complex
picture for the evolution of mixing of jet and external material.
Previous work suggests that significant toroidal magnetic fields can
reduce mass entrainment by suppressing filamentation of the jet beam.
Our present results suggest that the strong toroidal field component in
simulations C \& D has prevented the early development of elliptical
distortion seen in simulations C \& D from entraining much more mass
than is observed to be entrained in simulation B.  Nevertheless, it is
clear that the mass entrainment that occurs in simulations B, C \& D
and that is likely to occur in simulation A at larger spatial scales is
not conducive to highly collimated ``light'' jet propagation to distances 
orders of magnitude beyond the Alfv\'en point.

\section{Summary and Conclusion}

We have shown that jets remain K-H stable to low order asymmetric
normal modes, e.g., helical and elliptical modes, provided the jets are
on average sub-Alfv\'enic. Apparently, higher order normal modes, e.g.,
the rectangular surface mode, can be unstable provided a jet has a
sufficiently thick super-Alfv\'enic surface layer even when the jet is
on average sub-Alfv\'enic.  Jet rotation can be destabilizing when the
addition of the rotational velocity to the axial velocity gives a total
velocity shear parallel to the normal mode wavevector that is
super-Alfv\'enic. The lower order normal modes rapidly destabilize when
the jets become on average super-Alfv\'enic. We have observed
transverse velocity fluctuations associated with helical twisting at up
to twice the external sound speed, and half the jet's fast magnetosonic
speed in the super-Alfv\'enic and transmagnetosonic regime. Observed
jet distortions appear to be the result of K-H unstable helical,
elliptical, triangular, and rectangular normal surface modes.  No
evidence for the accompanying body modes or for the pinch normal mode
was evident in the simulations. These results are in agreement with a
linear stability analysis that indicates rapid growth rates for the
asymmetric normal surface modes on the super-Alfv\'enic and
transmagnetosonic jet, but that also indicates growth rates over an
order of magnitude lower for the accompanying body modes. We note that
the pinch surface mode is predicted to be K-H unstable on the
sub-Alfv\'enic and super-Alfv\'enic jet, albeit with relatively small
growth rate, and the pinch first body mode is predicted to be K-H
unstable on the super-Alfv\'enic jet with relatively high growth rate
but the present simulations provide no appropriate perturbation.

The primarily poloidally magnetized jet simulations (A \& B) indicate
that an initial small amplitude precessional motion is effectively
communicated down the jet to the Alfv\'en point.  This motion is
predicted to propagate at a wave speed $v_w\approx u+V_A$, although in
the present simulations the induced amplitude and transverse velocity
are too small, and the accompanying helical wavelength is too long to
directly observe the predicted wave motion and helical twist in the
sub-Alfv\'enic region.  The precessional perturbation couples to
growing jet distortions in the super-Alfv\'enic but transmagnetosonic
regime which propagate very slowly relative to the jet speed for the
``light'', $\rho _{jt}/\rho _{ex}\equiv \eta <<1$, poloidally
magnetized jets in simulations A \& B in accordance with a predicted
wave (pattern) speed of $v_w\approx [\eta /(1+\eta )]u$.

In simulations C \& D with a significant toroidal magnetic field
component, jet rotation accompanying the strong helical magnetic field
is on the order of the external sound speed (about 25\% of the
jet speed at the inlet), overwhelms the initial precessional motion by
over an order of magnitude, and the two simulations with opposite jet
precession develop nearly identically.  The jets in these simulations
destabilize rapidly to elliptical jet distortion and subsequently to
helical twisting of wavelength comparable to that in the poloidal
magnetic field simulations A \& B.  The elliptical and helical
distortions corotate with the jet rotation in accordance with
expectations based on a linear stability analysis of the poloidally
magnetized rotating jet (Bodo et al.\ 1996). The elliptical distortion
in these two simulations appears before the jet becomes
super-Alfv\'enic if only the axial jet velocity is considered.  The
presence of toroidal magnetic field appears to suppress the higher
order normal modes that on poloidally magnetized jets are rapidly
growing. This effect has also been observed in supermagnetosonic jet
simulations containing a significant toroidal magnetic field (RHCJ).

No significant mass entrainment occurs in the sub-Alfv\'enic region in
the poloidally magnetized simulations A \& B but significant mass
entrainment accompanies the development of helical twisting and
elliptical distortion in the super-Alfv\'enic region. The large
amplitude helical and elliptical jet distortions that accompany
destabilization are associated with a total (magnetic plus thermal)
pressure fluctuation at only the 10\% -- 20\% level. Comparison between
simulations and stability theory reveals that the distortions seen in
the simulations are at relatively long wavelengths relative to the
fastest growing wavelengths associated with the normal modes, and that
such ``long'' wavelength distortions can be induced by relatively small
total pressure fluctuations. Since variation in the synchrotron
emissivity is a function of the total pressure fluctuation, intensity
images, e.g., simulation A, can show very little apparent structure
even when jet cross sections and transverse velocity plots show readily
identifiable albeit relatively small amplitude structure.  Intensity
images do reveal evidence for helical twisting related to the
precession once the amplitude has become sufficiently large, e.g.,
simulation B, and some narrowing and broadening of the jet in the
intensity image can be identified with elliptical distortion of the jet
cross section. It is not possible to identify higher order jet
distortions seen in jet cross sections with features in the intensity
images.

In simulations C \& D jet rotation induced by the helical magnetic
field dominates the induced precessional motion at the inlet so in
spite of the opposite precession the two jets develop similarly. The
large jet rotation makes the jets unstable and effectively
super-Alfv\'enic almost from the inlet, and significant mass
entrainment begins earlier than would be expected based on the location
of the Alfv\'en point on the jet axis. An apparent oscillation in jet
width in the intensity image beginning near to the jet inlet is shown
by jet cross section to be the result of an elliptical distortion, and
not the result of a pinching distortion.  Intensity images indicate an
initial regular helical twist that develops into a complicated twisted
structure which terminates in a loop in the intensity images. Axial
velocity contours and transverse velocity vectors reveal a hollow jet
with large azimuthal motion around the intensity loop but not through
the loop axially as the loop moves outwards with the average outward
flow speed.  To our knowledge this is the first indication that jets
can become hollow and develop an azimuthal circulation leading to the
formation of a tight loop in an intensity image. Mass entrainment and
slowing of the jet outflow associated with the loop structure may
indicate the formation of a jet front.

The rapid onset of relatively short length scale jet distortion and
accompanying mass entrainment that we observe in these simulations when
they become super-Alfv\'enic but remain transmagnetosonic suggests the
development of a plume like appearance at larger distances than we can
simulate. Thus, the present simulations would be more appropriate to FR
I type extragalactic jets whose appearance has been argued, cf.,
Bicknell (1994, 1995) to be the result of significant mass entrainment.
The helically twisted structures that we do observe in the simulations
would be on much shorter spatial length scales than those of twisted
structures seen at parsec and kiloparsec length scales on extragalactic
jets, e.g., M87 (Reid et al. 1989; Junor \& Biretta 1995; Owen, Hardee
\& Cornwell 1989), unless helically twisted structures scale spatially
with, say, the jet radius beyond the Alfv\'en point. 

If magnetic jet acceleration and collimation schemes are to prove
viable for the production of observed protostellar, and, in particular,
FR II type extragalactic jets that  propagate to distances orders of
magnitude larger than the location of the Alfv\'en point, they must
flow through the transmagnetosonic region much more stably than the
jets in these simulations. We note that our present light jet
simulations with relatively flat internal density, velocity and
magnetic profiles are expected to be relatively unstable in both the
linear and non-linear growth regimes (cf. HCR, RHCJ). Additional
stability both linearly and non-linearly may be achieved by different
density, temperature, magnetic and velocity profiles, and through a
higher jet density relative to the surrounding environment, or by the
embedding of a jet in a surrounding wind. It is also possible that
relativistic effects associated with jet and Alfv\'en speeds near to
lightspeed will significantly modify the results and lead to a picture
more consistent with extragalactic jets. Only future work designed
to study relativistic effects, and to study the stability and mass
entrainment properties of jet profiles consistent with those predicted
to emerge from present magnetic acceleration and collimation schemes
will answer these questions.

\vspace {1.0cm}

P. Hardee and A. Rosen acknowledge support from the National Science
Foundation through grant AST-9318397 and AST-9802955 to the University of
Alabama. The authors would also like to acknowledge David Clarke who has
provided valuable support through development and maintenance of ZEUS-3D.
The numerical work utilized the Cray C90 at the Pittsburgh Supercomputing
Center through grant AST930010P.

\newpage 

\section{Figure Captions}

\figcaption{ 
Toroidal magnetic field profile (solid line) and thermal pressure
profiles in simulation A (dotted line), simulation B (long dashed line)
and simulations C \& D (short dashed line) at the inlet. Jet thermal
pressures are normalized to the pressure in the external medium at the
inlet and jet toroidal magnetic field is normalized to the peak
toroidal magnetic field.}

\figcaption{ 
Profiles of the velocity components scaled relative to the sound speed,
$a_{ex}$, in the external medium along the $z$-axis at dynamical times
$\tau _d=$ (A) 68, (B) 54, (C\ \&\ D) 44 in the four simulations from
simulation A (top row) to simulation D (bottom row).  The arrows indicate
the location of the Alfv\'en point on the jet axis.}

\figcaption{ 
Axial and transverse profiles in simulations A (top row) to D (bottom
row) at the dynamical times used in Figure 2. In the first column the
panels show values of the axial jet speed (solid line), jet sound speed
(dotted line), Alfv\'en speed (long dashed line), and fast magnetosonic
speed (dashed and dotted line) along the $z$-axis.  In column two the
panels show profiles of these speeds along the $x$-axis.  Profiles
along the $x$-axis are at locations $z = $ (A) $38.7R_0$, (B)
$18.1R_0$, (C) $14.0R_0$, and (D) $15.0R_0$ indicated by the vertical
lines in the panels in column one.  In columns three through five the
panels show profiles along the $x$-axis of: (col.3) density, $\rho$,
(solid line) and pressure, $e$, (dashed line), (col.4) temperature,
$e/\rho$, and (col.5) total magnetic field, $b$, (solid line) and
magnetic field components $b_z$ (dotted line), $b_y$ (dashed line),
respectively. The density, pressure, and temperature are scaled to the
density and 10/9 $\times$ the pressure in the external medium at the
inlet, and to 10/9 $\times$ the temperature in the external medium,
i.e., $\rho_{ex}(0)$, $1.11P_{ex}(0)$, and $1.11T_{ex}$, respectively.
The speeds are scaled relative to the sound speed, $a_{ex}$, in the
external medium.  The magnetic field strength is found from $B = 1.291
\surd P_{ex}(0) \times b$.  Note that the vertical scales are not all
identical.}

\figcaption{ 
Profiles of the average axial speed, $\left\langle
v_z\right\rangle $, of jet plus entrained mass, and of
the jet plus entrained mass per unit length, $\sigma /\sigma _{jt}$, in
simulation A (solid line), B (dashed \& dotted line), C (dotted line)
and D (dashed line).}

\figcaption{ 
Axial velocity cross sections where dark indicates high values at
axial distances from $6R_0$ to $60R_0$ in $6R_0$ increments in simulation A
(top) to simulation D (bottom). The $x$-axis is in the vertical direction, the
$y$-axis is in the horizontal direction and the flow direction ($z$-axis) is
into the page. Each cross section has a dimension $4R_0\times 4R_0$.}

\figcaption{ 
Intensity images of dimension $20R_0\times 70R_0$ with fractional
polarization B-vectors from simulations A (top) to D (bottom).}

\figcaption{ 
Solutions to the dispersion relation are shown as a function of angular
frequency at three locations along the jet in simulation B. Surface
pinch (P), helical (H), elliptical (E), triangular (T), and rectangular
(R) , and pinch 1st body (Pb$_1$) modes are shown. The real part of the
wavenumber, $k_R$, is indicated by the dotted lines and the absolute
value of the imaginary part of the wavenumber, $k_I$, is indicated by
the dashed lines. The precession frequency is $\omega R/u\approx 0.11$
at all three locations.}

\figcaption{ 
Displacement cross sections for helical, elliptical, triangular, and
rectangular surface waves appropriate to simulation B along with 1D
total pressure and velocity slices as a function of $z$ in
units of the jet radius, $R$. The 1D pressure and velocity slices are
taken at positions on the $y$-axis indicated by the ``$\times$'' in the
displacement cross sections. The dotted lines and the dashed lines indicate
the $x$ and $y$ components of the velocity, respectively. Note that
$v_y$ is a radial motion of the fluid and $v_x$ is an azimuthal motion
of the fluid. The radial velocity lags the azimuthal velocity in phase,
and the radial and azimuthal amplitudes can have different offsets
about zero velocity.}

\figcaption{ 
Total pressure, axial, and transverse velocity components from
simulation B as a function of $z$ in units of $R_0$. The top panels are
1D slices along the $z$-axis and the bottom panels are 1D slices
parallel to the $z$-axis at $x = 0.7R_0$. The dotted lines and the dashed
lines indicate the $x$ and $y$ components of the transverse velocity,
respectively. Note that these 1D slices are on the transverse axis
orthogonal to that used for the theoretical 1D slices shown in Figure
8. Now $v_x$ is a radial motion of the fluid and $v_y$ is an azimuthal
motion of the fluid. As in Figure 8 the radial velocity lags the azimuthal velocity in
phase.}

\figcaption{ 
Transverse slices showing transverse velocity vectors in
simulation C along the $z$-axis from $2R_0$ to $10R_0$ in $2R_0$ increments.
Typical rotation speed is on the order of $a_{ex}$. Each cross section
has a dimension $3R_0\times 3R_0$, and a vector is shown for every other
computational zone.}

\figcaption{
Contours of the axial velocity (first panel), transverse velocity
vectors (second panel), and contours of the flow angle relative to the
$z$-axis (third panel) in a transverse slice of dimension $8R_0\times
8R_0$ from simulation C at an axial distance of $42R_0$. Contours in
velocity are at intervals of $0.5a_{ex}$ and the bold arrow at the top
of the velocity vector panel has a length of $4a_{ex}$. Note the dashed
line contour that indicates negative (backwards) axial velocities on
the order of $0.5a_{ex}$. The largest transverse velocities are about
$5a_{ex}$.}

\newpage

\begin{table} 
 \begin{center}
 \caption{Initial Conditions \label{tbl-1}}
 \vspace{0.1cm}
 \begin{tabular}{ c c c c c c c c c} \hline \hline
 {\bf Simulation} & $a_{jt}/a_{ex}$ & $V_{A}/a_{ex}$ & $a_{ms}/a_{ex}$ 
                  & $\left\langle M_{jt}\right\rangle$ 
                  & $\left\langle M_{A}\right\rangle$ 
                  & $\left\langle M_{ms}\right\rangle$ 
                  & $C_p$ & $C_{\phi}$        \\ \hline
     A       &  2.32          &  6.00           & 6.43          &  1.85          &  0.66          &  0.62         &  6.567     &  0.050        \\
     B       &  3.71          &  5.00           & 6.23          &  1.08          &  0.80          &  0.64         &  1.526     &  0.004        \\
     C, D    &  5.40          &  5.00           & 7.36          &  1.01          &  0.77          &  0.61         &  1.526     &  0.298        \\  \hline
  \end{tabular}
  \end{center}

\end{table}

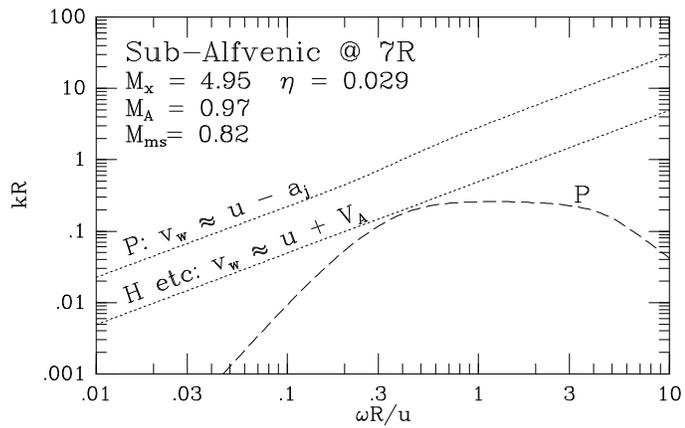
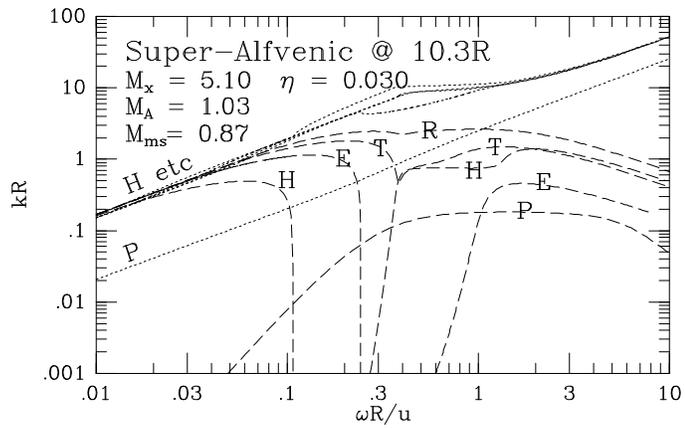
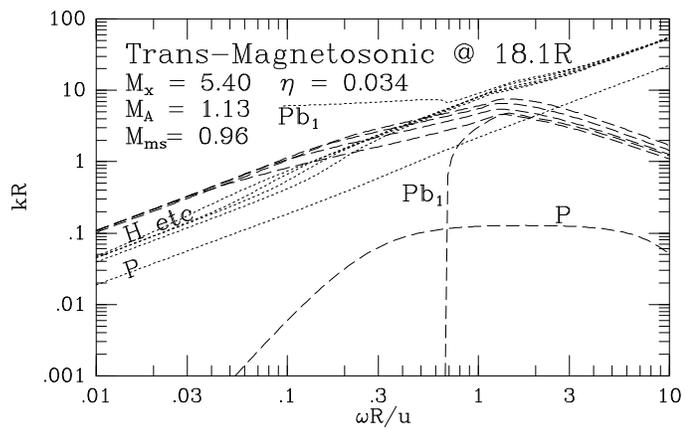
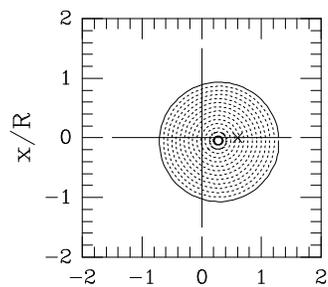
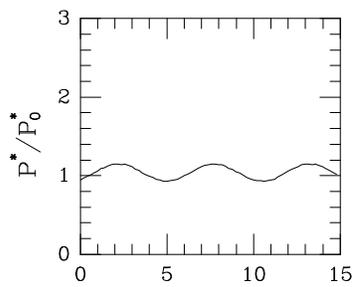
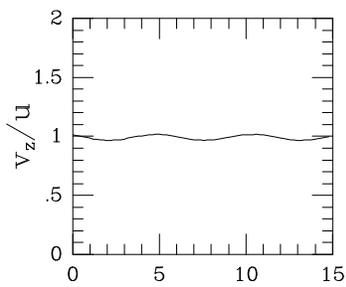
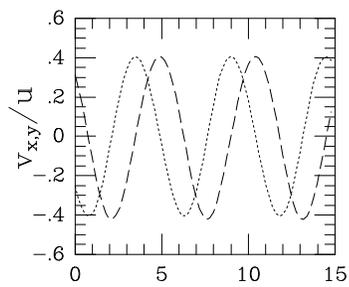
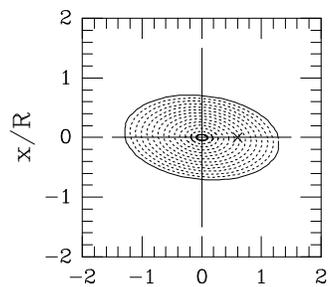
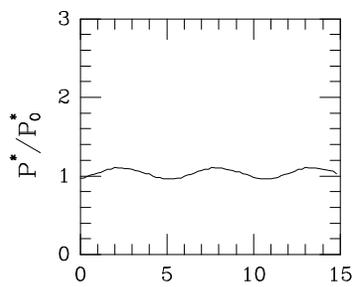
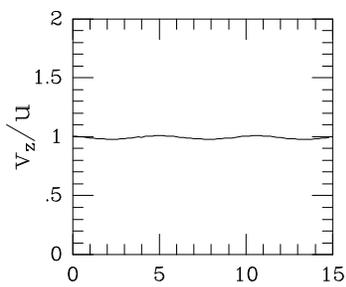
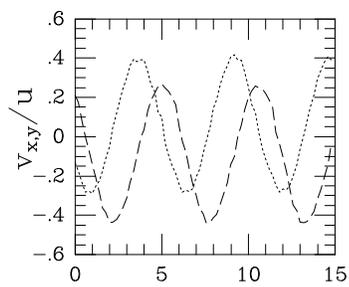
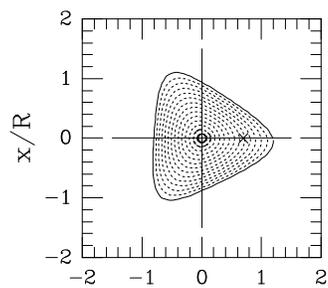
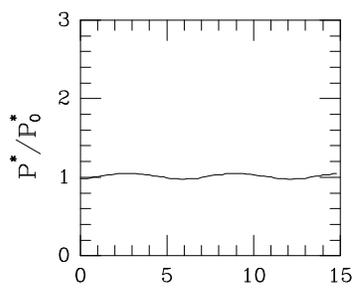
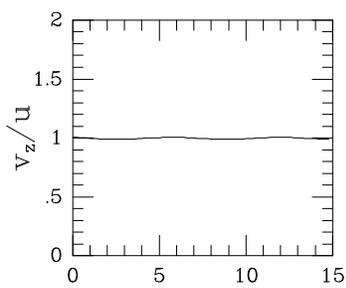
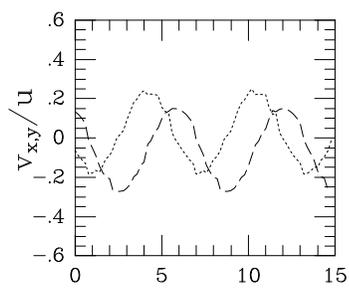
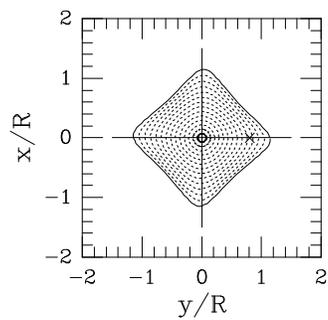
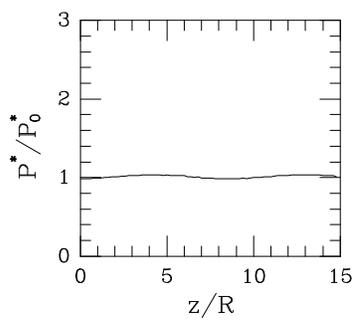
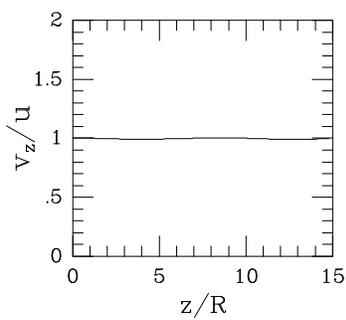
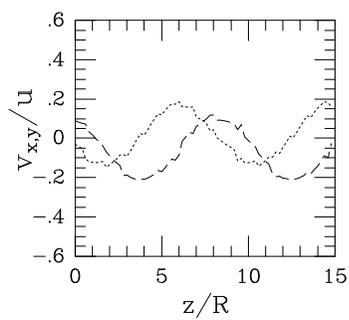
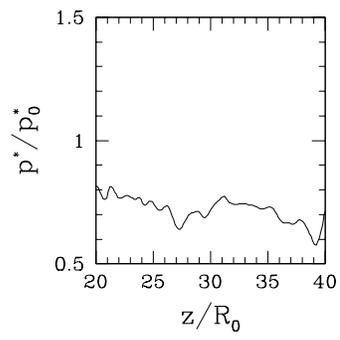
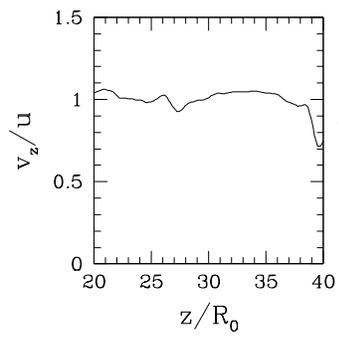
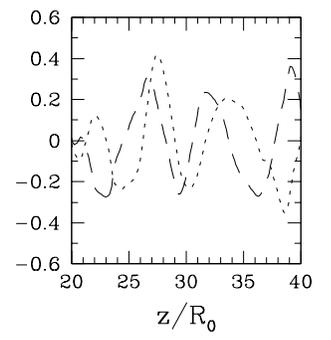
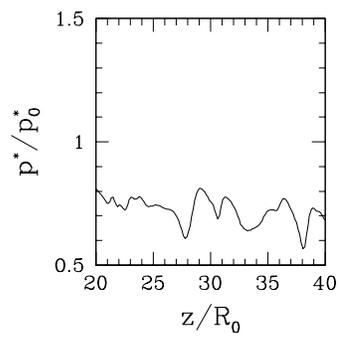
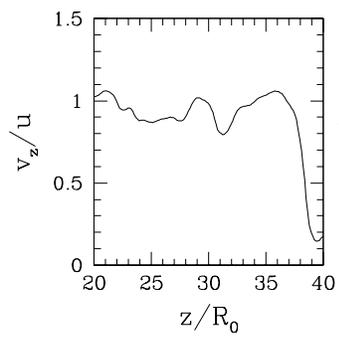
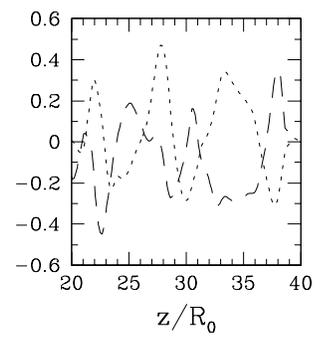
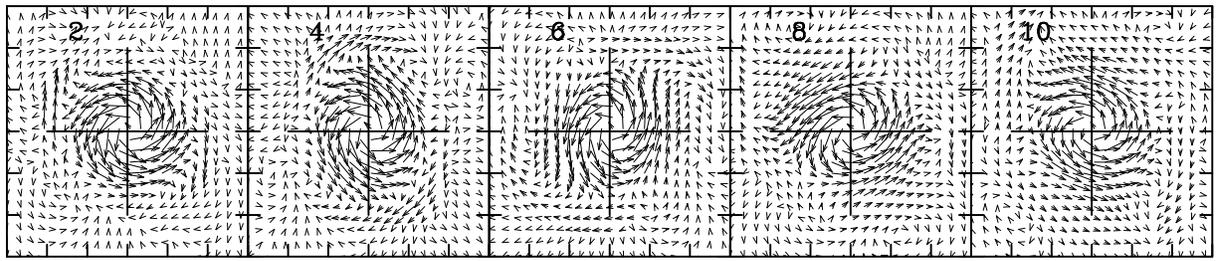
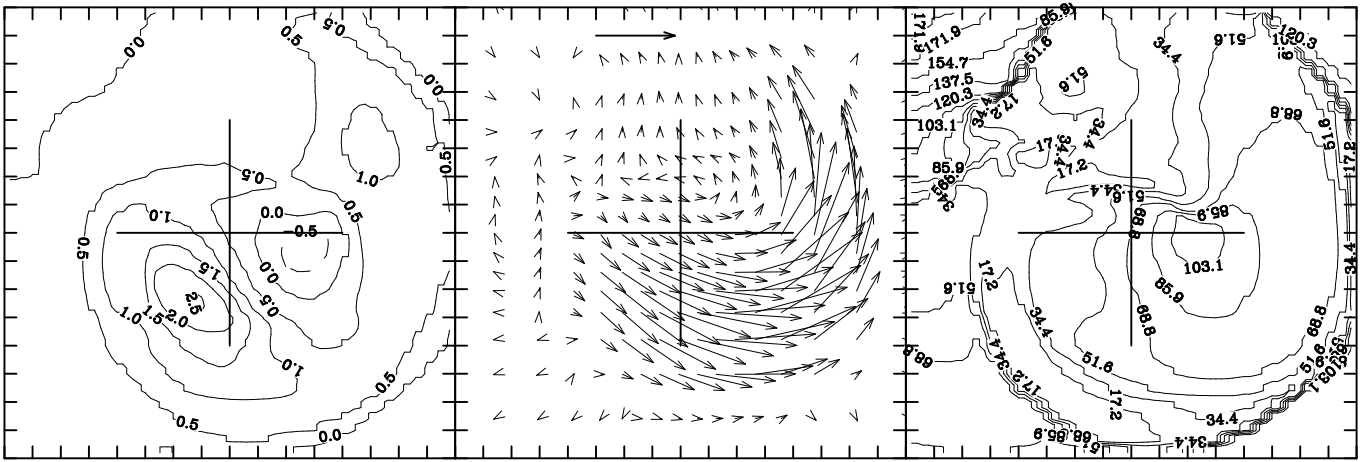
\end{document}